\documentclass[10pt]{iopart}
\usepackage{iopams}
\usepackage{graphicx}
\usepackage[usenames,dvipsnames]{color}
\input{epsf}
\newcommand{\trm}{\mathfrak{t}^{\Omega}}
\newcommand{\rfl}{\mathfrak{r}^{\Omega}}
\newcommand{\Renv}[3]{R^{#1}_{#2}(#3)}
\newcommand{\Lenv}[3]{L^{#1}_{#2}(#3)}
\newcommand{\Menv}[3]{M^{#1}_{#2}(#3)}
\newcommand{\Penv}[3]{P^{#1}_{#2}(#3)}

\newcommand{\Pq}{P^{\Omega}}
\newcommand{\Pan}{\langle P^{\Omega}\rangle}
\newcommand{\Pav}{P^{\langle \Omega\rangle}}
\newcommand{\Mav}{M^{\langle \Omega\rangle}}
\newcommand{\rav}{r^{\langle \Omega\rangle}}
\newcommand{\lav}{l^{\langle \Omega\rangle}}
\newcommand{\pav}{p^{\langle \Omega\rangle}}
\newcommand{\mav}{m^{\langle \Omega\rangle}}
\newcommand{\Rav}[2]{R^{\langle \Omega\rangle}_{#1}(#2)}
\newcommand{\Lav}[2]{L^{\langle \Omega\rangle}_{#1}(#2)}

\newcommand{\rflav}{\mathfrak{r}^{\langle\Omega\rangle}}

\begin{document}

\title{Transport properties and ageing for the averaged L\'evy-Lorentz gas}

\author{Mattia Radice$^{1,2}$, Manuele Onofri$^{1,2}$, Roberto Artuso$^{1,2}$, Giampaolo Cristadoro$^3$}
\address{$^1$ Dipartimento di Scienza e Alta Tecnologia and Center for Nonlinear and Complex Systems, Universit\`a degli Studi dell'Insubria, Via Valleggio 11, 22100 Como Italy}
\address{$^2$ I.N.F.N. Sezione di Milano, Via Celoria 16, 20133 Milano, Italy}
\address{$^3$ Dipartimento di Matematica e Applicazioni, Universit\`a degli Studi di Milano - Bicocca, Via  Cozzi 55, 20125 Milano, Italy}
\eads{\mailto{m.radice1@uninsubria.it}, \mailto{m.onofri1@uninsubria.it}, \\\mailto{roberto.artuso@uninsubria.it}, \mailto{giampaolo.cristadoro@unimib.it}}
\vspace{10pt}

\begin{abstract}
We consider a persistent random walk on an inhomogeneous environment where the reflection probability depends only on the distance from the origin. Such an environment  is the result of an average over all realizations of disorder of a L\'evy-Lorentz (LL) gas.  Here we show that this \emph{averaged L\'evy-Lorentz gas} yields nontrivial results even when the related LL gas is trivial. In particular, we investigate its long time transport properties such as the mean square displacement and the statistics of records, as well as the occurrence of ageing phenomena.
\end{abstract}

\vspace{2pc}

\section{Introduction}
The Lorentz gas was introduced in 1905 \cite{LG5} to model transport properties of electrons in metals: in its standard form it consists in a point particle moving in a periodic array of scatterers, with which it collides elastically. Translational symmetry allows to infer many properties of this extended system from the reduced dynamics in an elementary cell (see for instance \cite{book}). In particular dynamical and transport properties are deeply influenced by the shape of scatterers and the geometry of the lattice: circular obstacles lead to a Sinai billiard for the reduced dynamics, and the chaotic properties of such a system (see \cite{ChMa}) are crucial in dealing with the extended case. The geometry of the lattice, and eventually the size of the scatterers, determine whether particles may travel for arbitrarily long times without experiencing any collision: the so called infinite horizon case, which leads, in two dimensions, to a logarithmic correction to the variance of the particle position. For a recent review of (transport) properties of different types of Lorentz gases see \cite{dett}.

Much less is known about dynamical and transport properties when the scatterers are placed randomly, breaking translational symmetry \cite{MLap}. This motivates the introduction of simplified models, which however still present, as we will see, considerable complexity: a major role is played by persistent random walks in one dimension. Persistency consists in assigning to each site a transmission and a reflection coefficient in such a way that each step the walker undertakes at time $n$ not only depends on the position at the same time, but also on where the walker was one time step earlier. It is interesting to observe that (homogeneous) persistent random walks have been introduced nearly a century ago, as a model of diffusion by discontinuous movements \cite{Fu-p,Ta-p}: many of the relevant results have been obtained (for different time regimes) by introducing appropriate continuum limits \cite{Go-p,Ka-p,RH-p,We-p}. Inhomogeneous persistency arises naturally when we distribute randomly scatterers in the lattice (empty sites being given a null reflection coefficient): models of this kind have been introduced in \cite{ST-ll} as a variant of Sinai diffusion (see \cite{B:Hu}).

In recent years a great interest emerged for the case of a diluted distribution of scatterers, whose mutual distance is characterized by a heavy tailed distribution \cite{BFK-ll,BCV-ll,BCLL-ll,acor,blp}. This model, known in the literature as L\'evy-Lorentz gas (LL gas), is made particularly interesting by the fact that the long ballistic flights performed by the walker
are due to the
nature of the medium, not to a special law governing the walker's
decision, which is the case, e.g., of an homogeneous L\'evy walk \cite{ZDK,
	CGLS}. This produces a significant difference in the diffusion
properties: for example, a LL gas in an environment where the distance
between obstacles is L\'evy-distributed with parameter $\alpha \in
(1,2)$ is diffusive \cite{BCLL-ll}, but the L\'evy walk whose flights have the same
distribution is not \cite{CGLS, MSSZ}. Besides theoretical interest the LL-gas is tightly connected to experimentally fabricated L\'evy glasses \cite{BBW-ll}.
 
In order to study a mean field evolution over a fast changing environment we  have introduced in  \cite{acor} an \emph{averaged} landscape (constructed   by averaging   different environments). This construction leads  to a non-homogeneous persistent random walk where the reflection probability depends non-trivially on the distance from the origin. The subject of this paper is to introduce a variation of such an averaged model where we reverse the role of transmission and reflection with respect to the one defined in \cite{acor}, and to present some results regarding record statistics and ageing phenomena for both models. The paper is organized as follows: in section (\ref{s:2}) we provide the general setting, in the framework of persistent random walks, and the different probabilistic points of view that may be defined, then in section (\ref{s:3}) we introduce the ``localized'' model and discuss the different continuum limits, we also remark how the standard picture for homogeneous persistent walks becomes considerably more complex when translational invariance is broken. In sections (\ref{s:4}) and (\ref{s:5}) we present some numerical results concerning the statistics of maxima and the number of records for our system, and ageing phenomena; finally in section (\ref{s:6}) we present our conclusions.

\section{\label{s:2}L\'evy-Lorentz gas: the quenched, the annealed and the averaged}
We start with a general setting: a persistent 1-d random walk with position dependent transmission and reflection coefficients. In the canonical (homogeneous) persistent random walk scheme a particle starts moving from position $ x_0=0 $ in a random direction and then at each site it is reflected or transmitted according to a certain (constant) probability. In the LL gas scheme instead reflection may only occur at certain positions, where the {\it scatterers} are present. Scattering sites are placed at random positions in such a way that the relative distances are distributed according to a L\'evy-like probability distribution function (PDF), i.e. a distribution that decays with a heavy polynomial tail for large values of the argument:
\begin{equation}
\label{eq:pwr-law}
\mu(\xi)\sim \xi^{-(1+\alpha)},\quad 0<\alpha<2. 
\end{equation}
Notice that in this range the variance of the distance diverges, and in the restricted range $ 0<\alpha\leq1 $ also the average distance is infinite.

The set of the positions of the scatterers through which transmittance and reflection are assigned among the lattice is called the {\it environment}, namely
\begin{equation*}\label{eq:environment}
\Omega=\big\{ j\, |\, j\textrm{-th site is a scatterer}\big\}.
\end{equation*}
When defining the environment, one can set the origin in two different ways: in the {\it equilibrium} case (see \cite{BFK-ll}) it can be placed at any point of the lattice; in the {\it nonequilibrium} case instead it is always occupied by a scatterer and the process is conditioned to start with a scattering event. In the rest of the paper we will only deal with this case, since it is the closest to experimental realizations \cite{BCV-ll}.

Given a realization of the environment, transmittance and reflection among the lattice are assigned according to
\[
\begin{array}{r@{}l}
\rfl_j &{}=k\cdot\delta_j^{\Omega}\\
\trm_j &{}=1-\rfl_j
\end{array}
\]
where
\[
\delta_j^{\Omega}=\left\{
\begin{array}{ll}
1 & \textrm{if }j\in\Omega\\
0 & \textrm{if }j\not\in\Omega
\end{array}
\right.
\]
and $ 0<k<1 $, which in the literature is commonly taken equal to $ 1/2 $. Notice that in the nonequilibrium case we always have $ \delta_0^{\Omega}=1 $. Time  evolution is written in its simplest form once we introduce the quantities $\Renv{\Omega}{j}{n}$ and $\Lenv{\Omega}{j}{n}$, representing the probabilities of being at site $ j $ after $ n $ steps with right and left momentum, respectively. These probabilities evolve according to the Chapman-Kolmogorov equations:
\begin{equation}
\label{eq:fKolmRL}
\begin{array}{r@{}l}
\Renv{\Omega}{j}{n+1} &{}=\trm_j\cdot \Renv{\Omega}{j-1}{n}+ \rfl_j \cdot \Lenv{\Omega}{j+1}{n}\\
\Lenv{\Omega}{j}{n+1}&{}=\trm_j\cdot \Lenv{\Omega}{j+1}{n}+ \rfl_j \cdot \Renv{\Omega}{j-1}{n}
\end{array}
\end{equation}
with the initial conditions
\begin{equation}
\begin{array}{r@{}l}
\label{eq:initRL}
\Renv{\Omega}{0}{0} &{}=\Lenv{\Omega}{0}{0}=\frac{1}{2}\\
\Renv{\Omega}{j}{0} &{}=\Lenv{\Omega}{j}{0}=0\quad\forall j\neq 0.
\end{array}
\end{equation}
From (\ref{eq:fKolmRL}) and (\ref{eq:initRL}) one can derive the equations and the initial conditions for the probability of the displacement of the walker $ \Penv{\Omega}{j}{n}=\Renv{\Omega}{j}{n}+\Lenv{\Omega}{j}{n} $ and the probability current $ \Menv{\Omega}{j}{n}=\Renv{\Omega}{j}{n}-\Lenv{\Omega}{j}{n} $:
\begin{equation}
\label{eq:fKolmPM}
\begin{array}{r@{}l}
\Penv{\Omega}{j}{n+1} &{}=\Renv{\Omega}{j-1}{n}+ \Lenv{\Omega}{j+1}{n}\\
\Menv{\Omega}{j}{n+1}&{}=(\trm_j-\rfl_j)\cdot\left[ \Renv{\Omega}{j-1}{n}- \Lenv{\Omega}{j+1}{n}\right]
\end{array}
\end{equation}
with
\begin{equation}\label{eq:initPM}
\begin{array}{r@{}l}
\Penv{\Omega}{0}{0} &{}=1 \\
\Menv{\Omega}{0}{0} &{}=0.
\end{array}
\end{equation}

The function $ \Pq $, the PDF of the process for a typical fixed environment $ \Omega $, is the interesting quantity in the study of the {\it quenched} version of the LL gas. The transport properties of the system can then be derived through the evaluation of the mean square displacement (MSD) \cite{ST-ll,BCLL-ll}. Few results have been established hitherto for $ \alpha<1 $, while for $ \alpha>1 $ it has been proven the validity of the central limit theorem \cite{BCLL-ll}, but, interestingly, convergence of the moments still remains an open question.

In order to take into account all possible configurations of disorder, one can consider $ \Pan $, the average over all possible environments of the respective $ \Pq $. Such a quantity is known in the literature as {\it annealed law} \cite{BCLL-ll}, so we refer to this version of the model as {\it annealed LL gas}.
In this setting it is possible to compute analytically the asymptotic behaviour of the MSD, which in the nonequilibrium case grows linearly in time only for $ \alpha>3/2 $, while for smaller values of the exponent anomalous diffusion is predicted \cite{BCV-ll}. Such results have been obtained with a scaling hypothesis for the probability distribution of the walker, which is decomposed into a central part and a subleading term describing the behaviour at large distances and recently investigated using the single-big-jump principle \cite{VBB}:
\begin{equation}\label{eq:ann_dec}
\langle P^{\Omega}(x,t)\rangle = \frac{1}{\ell(t)}\mathcal{F}\left(\frac{x}{\ell(t)}\right)+\mathcal{H}(x,t).
\end{equation}
The correlation length $ \ell(t) $ is determined by using estimates according to the related resistance model treated in \cite{BGA}, which give
\begin{equation}\label{eq:corr_length}
\ell(t)\sim\left\{
\begin{array}{ll}
t^{\frac{1}{1+\alpha}} & 0<\alpha<1\\
t^{\frac{1}{2}} & 1\leq\alpha.
\end{array}
\right.
\end{equation}
It is also remarkable that, as for the quenched version, the CLT is valid in the restricted range $ \alpha > 1 $, even though for $ 1\leq\alpha<3/2 $ the system is superdiffusive. Regarding the case $ 0<\alpha<1 $, it has been recently proven \cite{blp} the validity of a generalized CLT for the finite-dimensional distributions of the continuous-time process, with the same scaling exponent $ 1/(1+\alpha) $ appearing in the correlation length characterizing the PDF.

In \cite{acor} we proposed the {\it averaged} version of the LL gas where we studied the process on an ``averaged" environment: given the distribution of the distances between scatterers - see eq. (\ref{eq:pwr-law}) - one can compute the asymptotic form of the probability of finding a scatterer at site $ j $, which we denote as $ \pi_j $. This has the expression
\begin{equation}\label{eq:p_j}
\pi_j = \left\{
\begin{array}{ll}
\frac{\alpha\sin(\pi\alpha)}{\pi}\frac{\zeta(1+\alpha)}{|j|^{1-\alpha}} & 0<\alpha<1\\
\frac{\zeta(1+\alpha)}{\zeta(\alpha)} & 1<\alpha
\end{array}
\right.
\end{equation}
for $ j\neq0 $ ($ \zeta(s) $ is the Riemann zeta function), while $ \pi_0=1 $ since we are dealing with the nonequilibrium case (further details can be found in \cite{acor}). Therefore the average of the reflection coefficient at site $ j $ over all possible realizations of the environments can be evaluated as
\begin{equation}\label{eq:rflav}
\rflav_j=\frac 12\pi_j
\end{equation}
where we made the choice $ k=1/2 $, accordingly to the literature. Note that in the averaged environment every site is occupied by a scatterer, i.e. $ \Omega=\mathbb{Z} $, but the reflection coefficients among the lattice are taken equal to eq. (\ref{eq:rflav}). We report for completeness the main results: the system displays normal diffusion in the range $ 1<\alpha<2 $, while for $ 0<\alpha<1 $ the whole moments spectrum can be described as follows:
\begin{equation}\label{eq:av_MSD}
\langle|x_n|^q\rangle\sim n^{\frac{q}{1+\alpha}}
\end{equation}
which indicates that the model is weakly anomalous \cite{CMMV-d}. Indeed for both regimes there is a single scale ruling the whole moments spectrum, which is equal to the correlation length evaluated in the annealed model - see eq. (\ref{eq:corr_length}).

In this paper we will show that the averaged model gives nontrivial results even when the related quenched model is instead trivial. Indeed we consider a variation of the LL gas in which every empty site is filled with a perfectly reflecting barrier. The quenched version of this model - which we will call the {\it localized} model - is trivial, since the particle will be confined between two barriers and there will be no diffusion at all. The averaged version instead  keeps track of all the possible configurations of disorder: considering also the probability of being reflected at an empty site, the resulting average value of the reflection coefficient is
\begin{equation}\label{eq:rflav_loc}
\rflav_j=\frac{1}{2}\pi_j+(1-\pi_j)=1-\frac 12\pi_j.
\end{equation}
We will prove in the next section that for such a set of reflection coefficients the transport is not trivial and the system displays subdiffusion.

\section{\label{s:3}Continuum limits}

\subsection{The diffusion approximation for the averaged environment}
The continuum limit for a persistent random walk on a lattice has been considered in many of the classical papers, as \cite{Go-p, Ka-p, RH-p, We-p}. We call $ \delta x $ the lattice spacing, $ \delta t $ the time step and set $ x=j\delta x $ and $ t=n\delta t $. We use the short-hand notation $ \rav=r(x,t;\langle\Omega\rangle) $ and $ \lav=l(x,t;\langle\Omega\rangle) $ to denote the probability densities of being at position $ x $ at time $ t $ and leaving to the right or left respectively on the averaged environment, and write
\begin{equation}\label{eq:cont_RL}
\begin{array}{r@{}l}
\Rav{j}{n} &{}=\delta x\cdot \rav\\
\Lav{j}{n} &{}= \delta x\cdot \lav
\end{array}
\end{equation}
while for the quantities $ \Pav=R^{\langle\Omega\rangle}+L^{\langle\Omega\rangle} $ and $ \Mav=R^{\langle\Omega\rangle}-L^{\langle\Omega\rangle} $ we set
\begin{equation}\label{eq:cont_PM}
\begin{array}{r@{}l}
\Pav_j(n) &{}=\delta x\cdot \pav=\delta x\cdot p(x,t;\langle\Omega\rangle)\\
\Mav_j(n) &{}=\delta t\cdot \mav=\delta t\cdot m(x,t;\langle\Omega\rangle)
\end{array}
\end{equation}
where $ \pav $ and $ \mav $ are defined in terms of $ \rav $ and $ \lav $ as \cite{acor}:
\begin{equation}
\label{eq:def_pm}
\begin{array}{r@{}l}
\pav &{}=\rav+\lav\\
\mav &{}= \frac{\delta x}{\delta t}\cdot\left(\rav-\lav\right).
\end{array}
\end{equation}
Inserting (\ref{eq:cont_RL}) and (\ref{eq:cont_PM})  into the Chapman-Kolmogorov equations (\ref{eq:fKolmPM}) and expanding the functions $ \rav$, $\lav$, $\pav$, $\mav $ up to second order in both $ \delta x $ and $ \delta t $, one gets the following pair of coupled equations (we drop the superscript $ \langle\Omega\rangle $ for the functions and the coefficients):
\begin{equation}
\label{eq:coup_pm}
\left\{
\begin{array}{r@{}l}
\dot{p}\delta t+\frac{1}{2}\ddot{p}\delta t^2 &{}=-m'\delta t+\frac{1}{2}p''\delta x^2\\
m\delta t +\dot{m}\delta t^2+\frac{1}{2}\ddot{m}\delta t^3&{}=\left(\mathfrak{t}-\mathfrak{r}\right)\cdot\left(m\delta t-p'\delta x^2+\frac{1}{2}m''\delta x^2\delta t\right)
\end{array}
\right.
\end{equation}
where in the definition of $ \mathfrak{r} $ and $ \mathfrak{t} $ we use the continuous-space limit of (\ref{eq:p_j}):
\begin{equation}\label{eq:continuum_pi}
\pi_j\to\pi(x)=\left\{
\begin{array}{ll}
\frac{\alpha\sin(\pi\alpha)}{\pi}\frac{\zeta(1+\alpha)}{|x|^{1-\alpha}} & 0<\alpha<1\\
\frac{\zeta(1+\alpha)}{\zeta(\alpha)} & 1<\alpha.
\end{array}
\right.
\end{equation}

We get a closed equation for $ p $ by employing the diffusion approximation: we consider the limit $ \delta x,\delta t \to 0 $ with $ \delta x^2/\delta t=\Delta $ kept constant. By dropping higher order terms we obtain the following set of equations:
\begin{equation}\left\{
\begin{array}{r@{}l}
\frac{\partial p}{\partial t}&{}=-\frac{\partial m}{\partial x}+\frac{\Delta}{2}\frac{\partial^2 p}{\partial x^2}\\
m&{}=-\frac{\Delta}{2}\frac{\mathfrak{t}-\mathfrak{r}}{\mathfrak{r}}\frac{\partial p}{\partial x}
\end{array}
\right.
\end{equation}
Inserting the second one into the first, we finally get
\begin{equation}
\label{eq:diffusion}
\frac{\partial p}{\partial t}=\frac{\partial}{\partial x}\left[D_{\alpha}(x)\frac{\partial p}{\partial x}\right]
\end{equation}
where $ D_{\alpha}(x) $ is
\begin{equation}\label{eq:av_cont_diff_coeff}
D_{\alpha}(x)=\frac{\Delta}{2}\frac{\mathfrak{t}(x)}{\mathfrak{r}(x)}
\end{equation}
and we set $ \Delta=1 $. The explicit dependence of $ D_{\alpha} (x)$ on space is thus given through the definition of the reflection coefficient, which is displayed in equations (\ref{eq:rflav}) - averaged LL gas - and (\ref{eq:rflav_loc}) - averaged-localized LL gas. For both models $ D_\alpha(x) $ behaves like a power-law in the range $ \alpha\in\left(0,1\right) $ - with different powers for the two models - and it is constant for $ \alpha\in\left(1,2\right). $ Nevertheless, the derivation of the diffusion equation (\ref{eq:diffusion}) is independent on the definition of the reflection coefficient.

We point out that the eq. (\ref{eq:diffusion}) suggests that if one wishes to formalise the corresponding paths in the continuum limit, the corresponding Langevin equation should follow the H\"anggi-Klimontovich interpretation, rather than Stratonovich or It\^o. For a discussion about the effects of different interpretations on the solution's shape in the context of heterogeneous diffusion, as well as their role in infinite ergodic theory, see \cite{LB}.

\subsection{Solutions to the diffusion equation and evaluation of the mean square displacement}
Given the form of the diffusion coefficient (\ref{eq:av_cont_diff_coeff}), the diffusion equation (\ref{eq:diffusion}) can be solved exactly. For the case $ \rav_j=\frac 12 \pi_j $ the explicit solution is derived in \cite{acor}. Here we recall that in the regime $ \alpha > 1 $ - where, for the quenched LL gas, the average distance between scatterers is finite and the central limit theorem holds \cite{BCLL-ll} - we have a constant diffusion coefficient written as the ratio of transmission and reflection, in agreement with the result deduced in \cite{RH-p}.
In the range $ 0<\alpha<1 $ instead the diffusion coefficient is space-dependent, with a power-law of the form $ D_{\alpha}\sim|x|^{1-\alpha} $.
The behaviour of the whole moments spectrum is ruled by a single scale, therefore we have
\begin{equation}\label{eq:av_anMSD}
\langle|x_t|^q\rangle\sim \ell(t)^q=t^{\frac{q}{1+\alpha}}.
\end{equation}

We now analyse the localized case: the form of the reflection coefficient along the lattice is
\begin{equation}\label{key}
\rflav_j=1-\frac{1}{2}\pi_j
\end{equation}
and by using the continuum limit, the diffusion coefficient shows once again two different behaviours, depending on the value of $ \alpha $:
\begin{equation}\label{eq:inv_diff_const}
D_{\alpha}(x)=\left\{
\begin{array}{ll}
\frac{1}{2}\frac{1}{2\Lambda|x|^{1-\alpha}-1} & 0<\alpha<1\\
\frac{1}{2}\frac{1}{2\zeta(\alpha)/\zeta(1+\alpha)-1} & 1<\alpha<2
\end{array}
\right.
\end{equation}
where the constant $ \Lambda $ is
\begin{equation}\label{eq:Lambda}
\Lambda = \frac{\pi}{\alpha\sin(\pi\alpha)\zeta(1+\alpha)}.
\end{equation}

We remark that when $ \alpha\in\left(0,1\right) $, $ D_\alpha(x) $ as defined in eq. (\ref{eq:inv_diff_const}) can attain negative values in the region
\begin{equation}\label{key}
|x|<|x_c|=\left(2\Lambda\right)^{\frac{1}{\alpha-1}.}
\end{equation}
This is due to the fact that the limit we used is meaningful only in the region where the definition of $ \pi_j $ can be extended to continuum space, i.e. where $ \pi(x) $ is well-defined - see eq. (\ref{eq:continuum_pi}). Therefore to describe the process with continuum paths one would need to adopt some regularization of the diffusion coefficient inside the critical region - see the treatment of a similar case in \cite{LB}. However, since we are interested in the long-time properties, we don't expect this fact to affect the results.

\begin{figure*}[h!]
\centering
\begin{tabular}{c @{\quad} c }
\includegraphics[width=.45\linewidth]{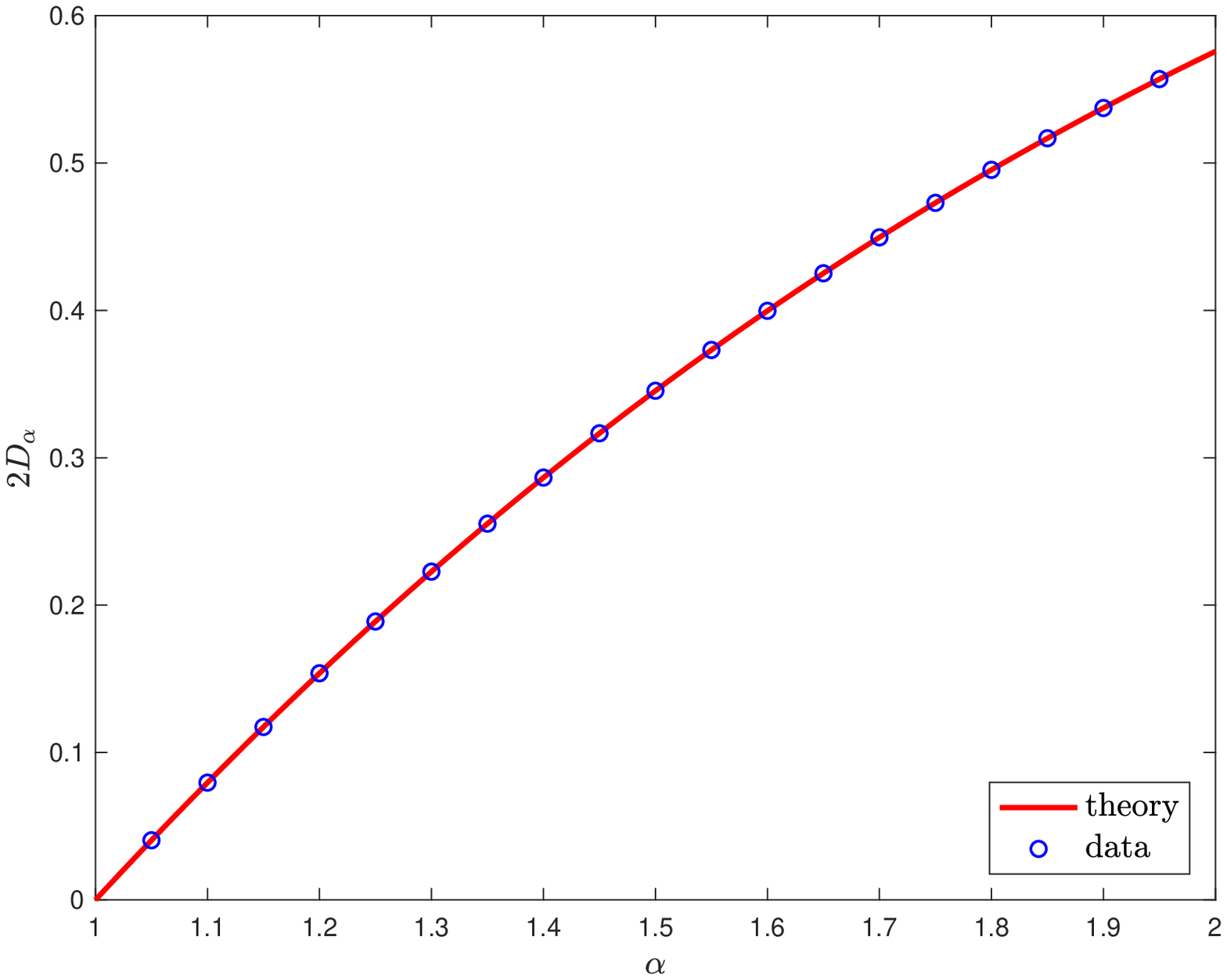} &
\includegraphics[width=.45\linewidth]{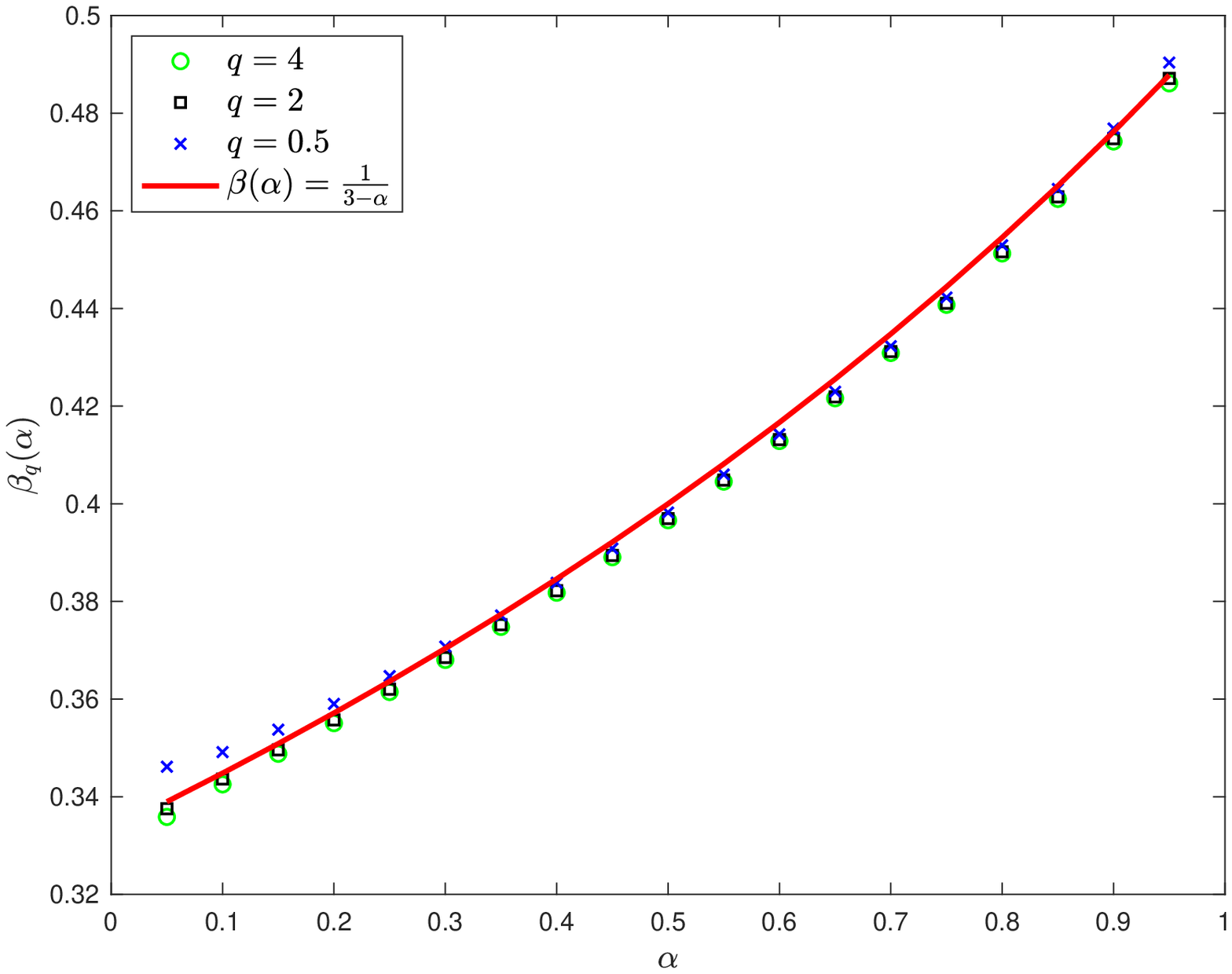} \\
\small (a) & \small (b)
\end{tabular}
\caption{Left: slope of linear growth of the second moment as obtained by numerically evolving the forward Kolmogorov equations (\ref{eq:fKolmPM}) (circles)  and the analytic prediction in terms of the diffusion constant (\ref{eq:inv_diff_const}). Each numerical slope has been obtained by evolving the system for $n=2^{15}$ number of steps. Right: asymptotic growth exponents of $ q $th order moments $ \langle|x_t|^q\rangle \sim t^{q\cdot\beta(\alpha)}$, for $ q=0.5 $ (blue crosses), $q=2$ (squares) and $q=4$ (green circles). Data are obtained evolving the system up to $ n=2^{18} $ number of steps.}
\label{fig:LL-diff}
\end{figure*}

We now pass to the solution of the diffusion equation and the evaluation of the MSD: for $ 1<\alpha<2 $, when the mean distance between scatterers is finite, $ D_{\alpha}(x) $ is constant and we have normal diffusion. In particular the MSD can be evaluated as follows:
\begin{equation}\label{key}
\langle x_t^2\rangle = \frac{t}{2\zeta(\alpha)/\zeta(1+\alpha)-1}
\end{equation}
in excellent agreement with numerical simulations (fig. \ref{fig:LL-diff}a).

When $ 0<\alpha<1 $ the diffusion coefficient is $ D_{\alpha}(x)\sim \Sigma|x|^{\alpha-1} $, with $ \Sigma=1/4\Lambda $; the solution of the diffusion equation reads \cite{RGF-d}
\begin{equation}\label{eq:loc_PDF}
p(x,t;\langle\Omega\rangle)=\frac{\left[(3-\alpha)^{\alpha-1}\Sigma t\right]^{-1/(3-\alpha)}}{2\Gamma(1/(3-\alpha))}\exp\left[-\frac{|x|^{3-\alpha}}{(3-\alpha)^2\Sigma t}\right].
\end{equation}
The system displays once again weak anomalous diffusion \cite{CMMV-d}, with the moments spectrum behaving as
\begin{equation}\label{key:exp:moments}
\langle|x_t|^q\rangle \sim t^{q\cdot\beta(\alpha)}
\end{equation}
where $ \beta(\alpha)=1/(3-\alpha) $ is the scaling exponent, such that the characteristic length of the distribution (\ref{eq:loc_PDF}) is
\begin{equation}\label{key}
\ell(t)=t^{\beta(\alpha)}
\end{equation}
as displayed by numerical simulations (fig. \ref{fig:LL-diff}b). In particular the MSD grows as
\begin{equation}\label{key}
\langle x_t^2\rangle \sim t^{\frac{2}{3-\alpha}}
\end{equation}
so that transport is subdiffusive.

\begin{figure*}[h!]
\centering
\begin{tabular}{c @{\quad} c }
\includegraphics[width=.45\linewidth]{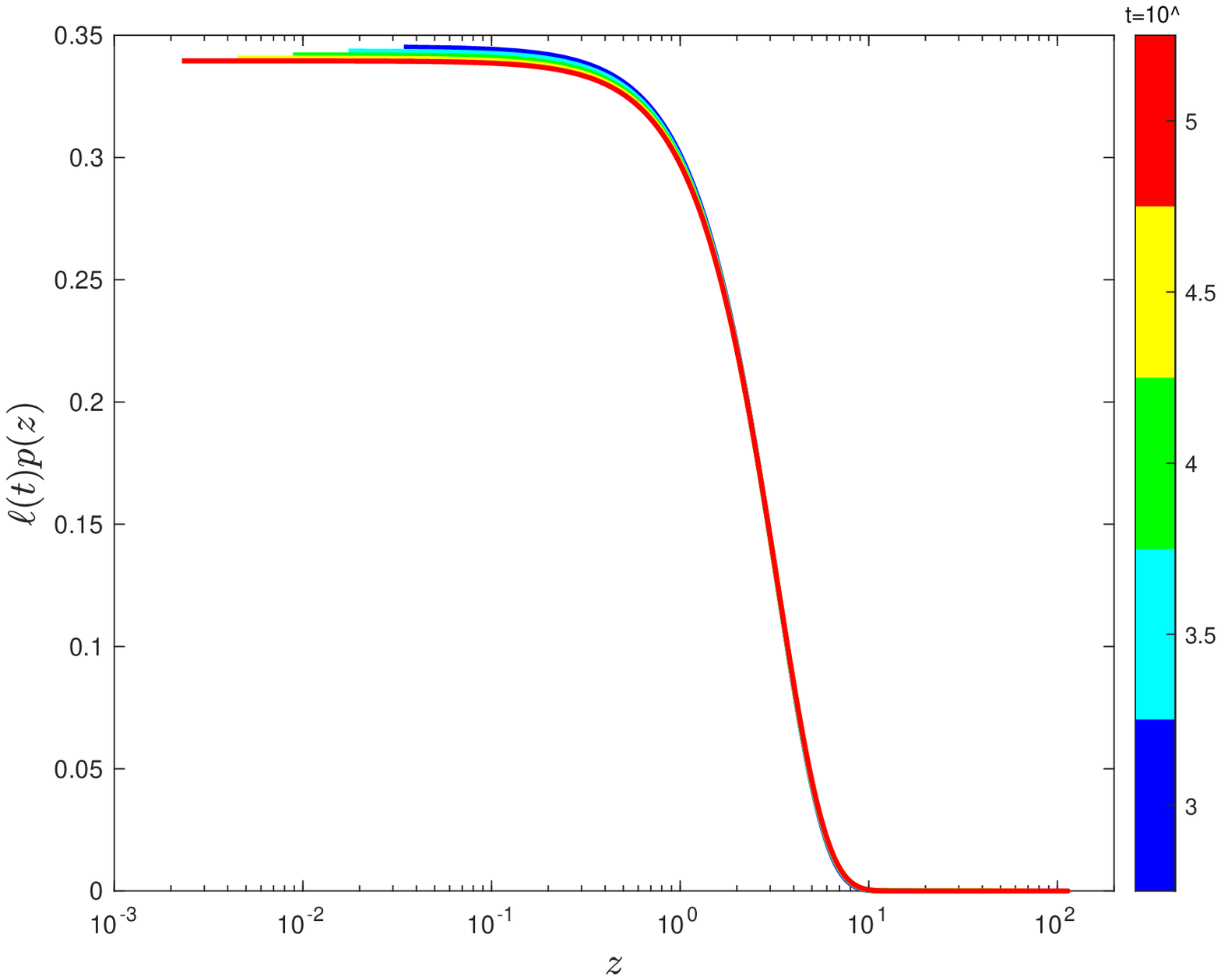} &
\includegraphics[width=.45\linewidth]{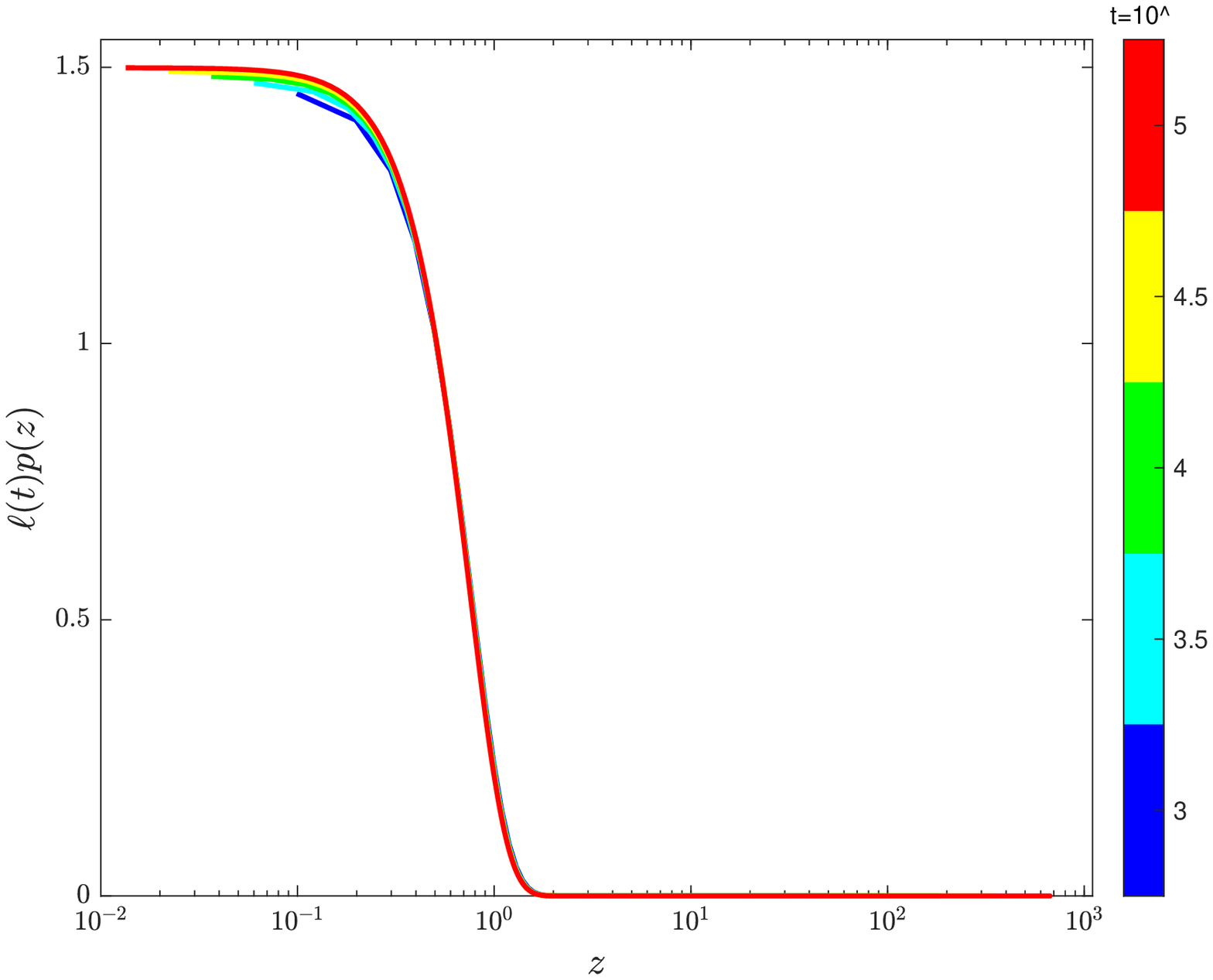} \\
\small (a) & \small (b)
\end{tabular}
\caption{Plot of the rescaled PDF $ \ell(t)p(z) $ versus the variable $ z=x/\ell(t) $, for positive values of $ x $, in logarithmic scale. On the left we present the superdiffusive model and on the right the subdiffusive one, both with $ \alpha=0.7 $. The pdf's are drawn letting evolve the system for different times up to $ t=10^5 $.}
\label{Cont_lim:scal_pdf}
\end{figure*}

We point out that the origin of the weak anomalous behaviour of the moments in our system can be traced back to the scaling form of the PDF. For both models the PDF in the continuum limit has the asymptotic form
\begin{equation}\label{key}
p(x,t;\langle \Omega\rangle)=\frac {1}{\ell (t)}\mathcal{F}\left(\frac{x}{\ell(t)}\right)
\end{equation}
as it is shown in fig. \ref{Cont_lim:scal_pdf} for a particular value of $ \alpha $.
Moreover, the fact that a single scaling length rules the whole moments spectrum enables us to derive informations on quantities related to the behaviour of the moments, as we will show in the next section.

\section{\label{s:4}Records statistics and expected maximum}
The statistics of records have become a subject of great interest in recent years, since it has applications in a large variety of fields, from meteorology \cite{Hoyt, Redner-Petersen}, to economics \cite{Barlevy}, to sports \cite{Gem-Taylor-Sut,Ben-Redner-Vazquez}: for reviews in the context of 1-d random processes see \cite{Maj, God-Maj-Sch}.

In a sequence $\left\lbrace x_0, x_1, \cdots ,x_n \right\rbrace$ the event $x_i$ is a called a {\it record} if it exceeds the values of all the previous data. In particular we are interested in the expected value of the maximum and the mean number of records for a random walk composed by $n$ steps performed in the averaged L\'evy-Lorentz gas, where the event $x_i$ is the position reached by the walker after $i$ steps. Since the motion occurs on a lattice with unit spacing and the particle starts the walk at $x_0=0$, which by definition is the first record, the number of records $\mathcal{N}_n$ after $n$ steps is equal to $ X_n+1 $, where $ X_n $ is the maximum value reached up to step $ n $. Therefore the expected maximum $\langle X_n \rangle$ and the mean number of records $\langle \mathcal{N}_n \rangle$  have the same asymptotic behaviour:
\begin{equation}
\langle \mathcal{N}_n \rangle \sim \langle X_n \rangle.
\end{equation}

\begin{figure*}[h!]
\centering
\begin{tabular}{c @{\quad} c }
\includegraphics[width=.45\linewidth]{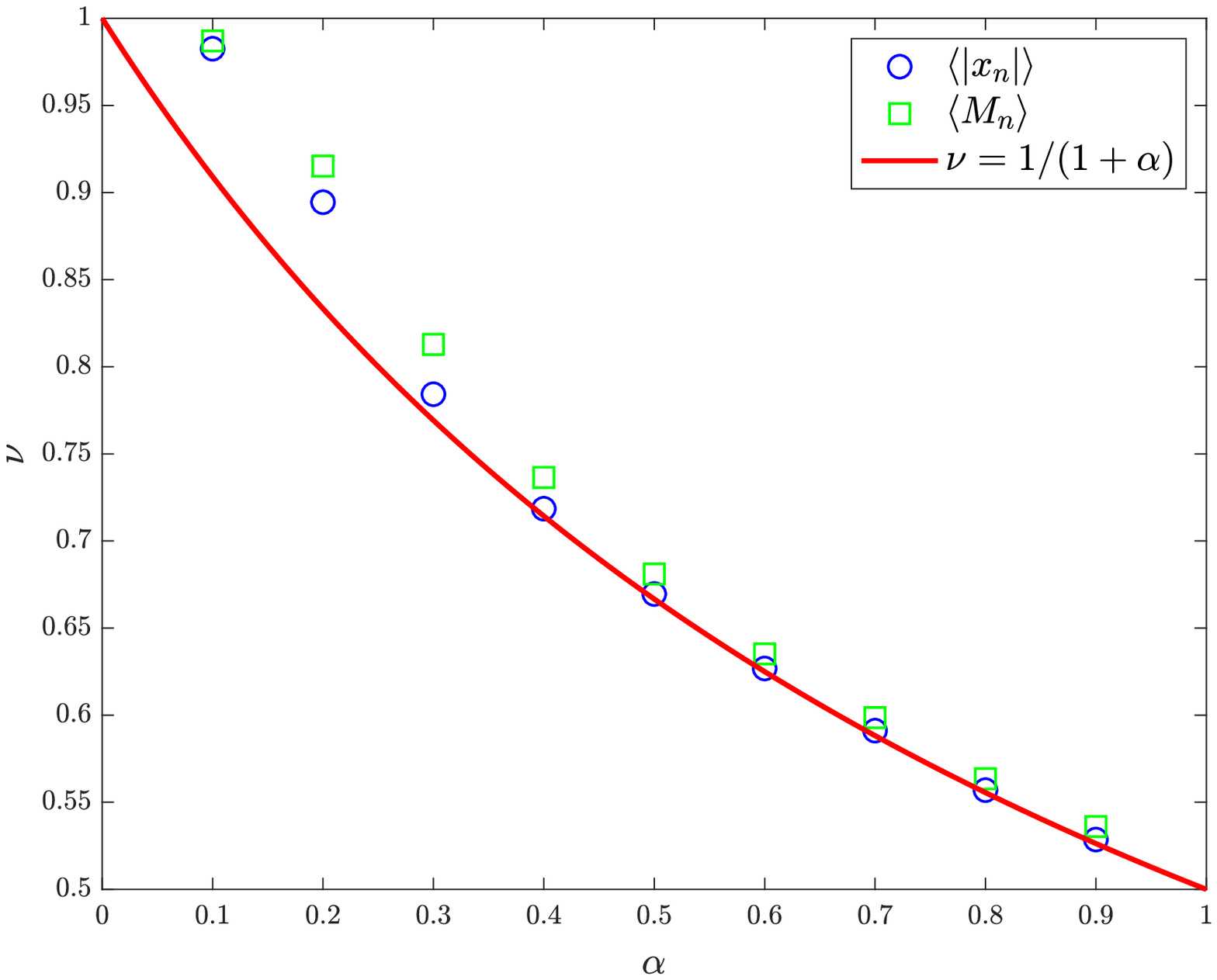} &
\includegraphics[width=.45\linewidth]{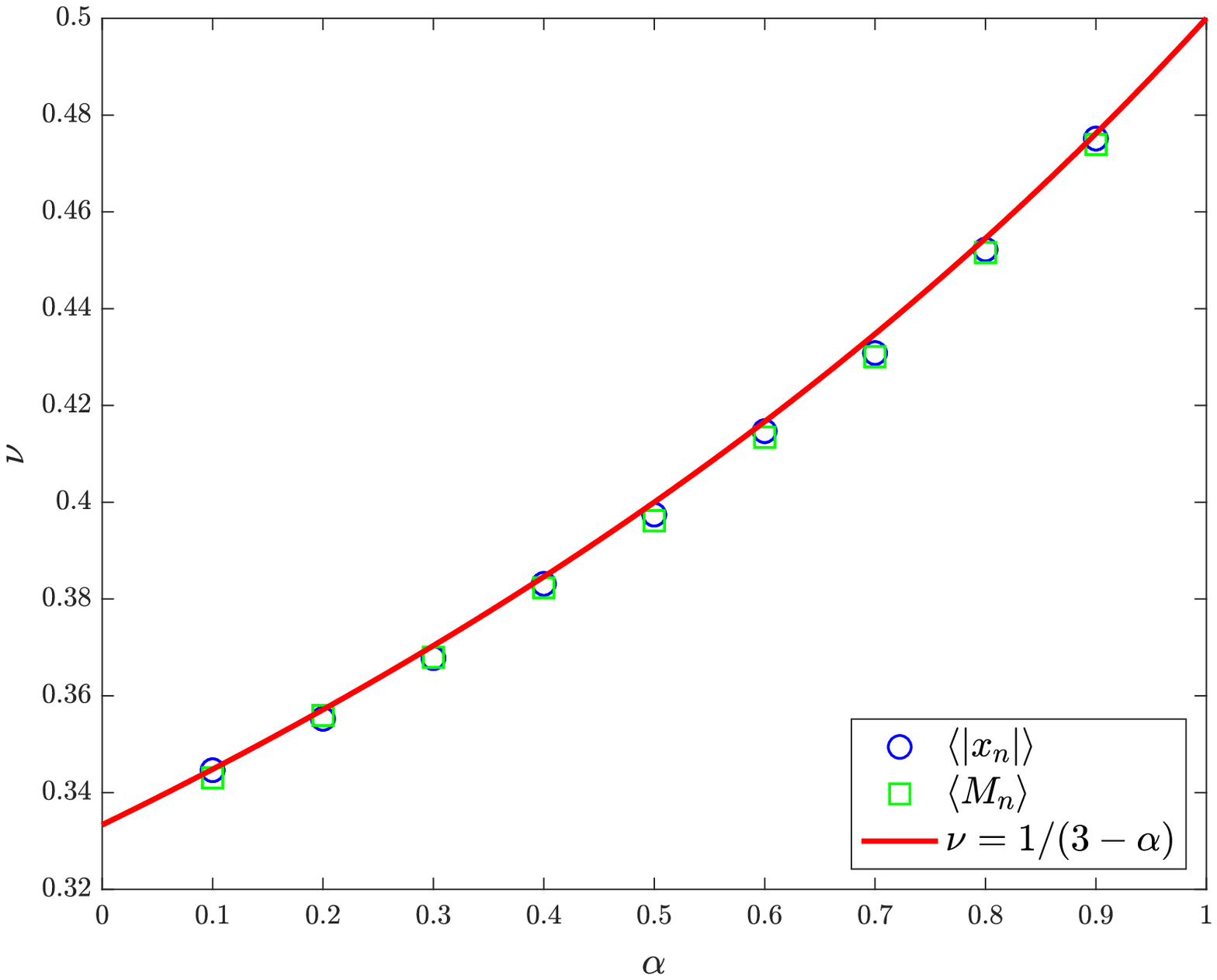} \\
\small (a) & \small (b)
\end{tabular}
\caption{Asymptotic growth exponents of the first moment $ \langle|x_n|\rangle$ and the expected maximum $\langle X_n\rangle$ depending on $\alpha$ both for the superdiffusive, on the left, and subdiffusive model, on the right. Data are obtained by averaging $N=10^6$ walks composed of $n=10^5$ number of steps.}
\label{Exp:record:first_moment}
\end{figure*}

For a random walk with independent jumps length drawn from an arbitrary symmetric density function the expected maximum grows asymptotically as the absolute first moment $\langle| x_n |\rangle$  \cite{Com-Maj}:
\begin{equation}\label{key}
\langle X_n\rangle\sim\langle|x_n|\rangle.
\end{equation}
No results are known for random walks with correlated jumps or space-dependent transition probabilities, so we tested numerically if a similar relation holds. As we can see in fig. \ref{Exp:record:first_moment} our guess is confirmed by simulations for the subdiffusive model (fig. \ref{Exp:record:first_moment}a) for all values of $\alpha$, while for the superdiffusive one (fig. \ref{Exp:record:first_moment}b) we report disagreement only for small values of $\alpha$. However, we remark that for the same values of $\alpha$ we also noticed in \cite{acor} that discrepancies with respect to the predictions of the diffusion approximation arose in the numerical evaluation of the moments, meaning that the diffusive regime is not yet captured at the number of steps of our simulations. We thus expect that a similar explanation is valid for the currently observed discrepancies.

\begin{figure*}[h!]
\centering
\begin{tabular}{c @{\quad} c }
\includegraphics[width=.45\linewidth]{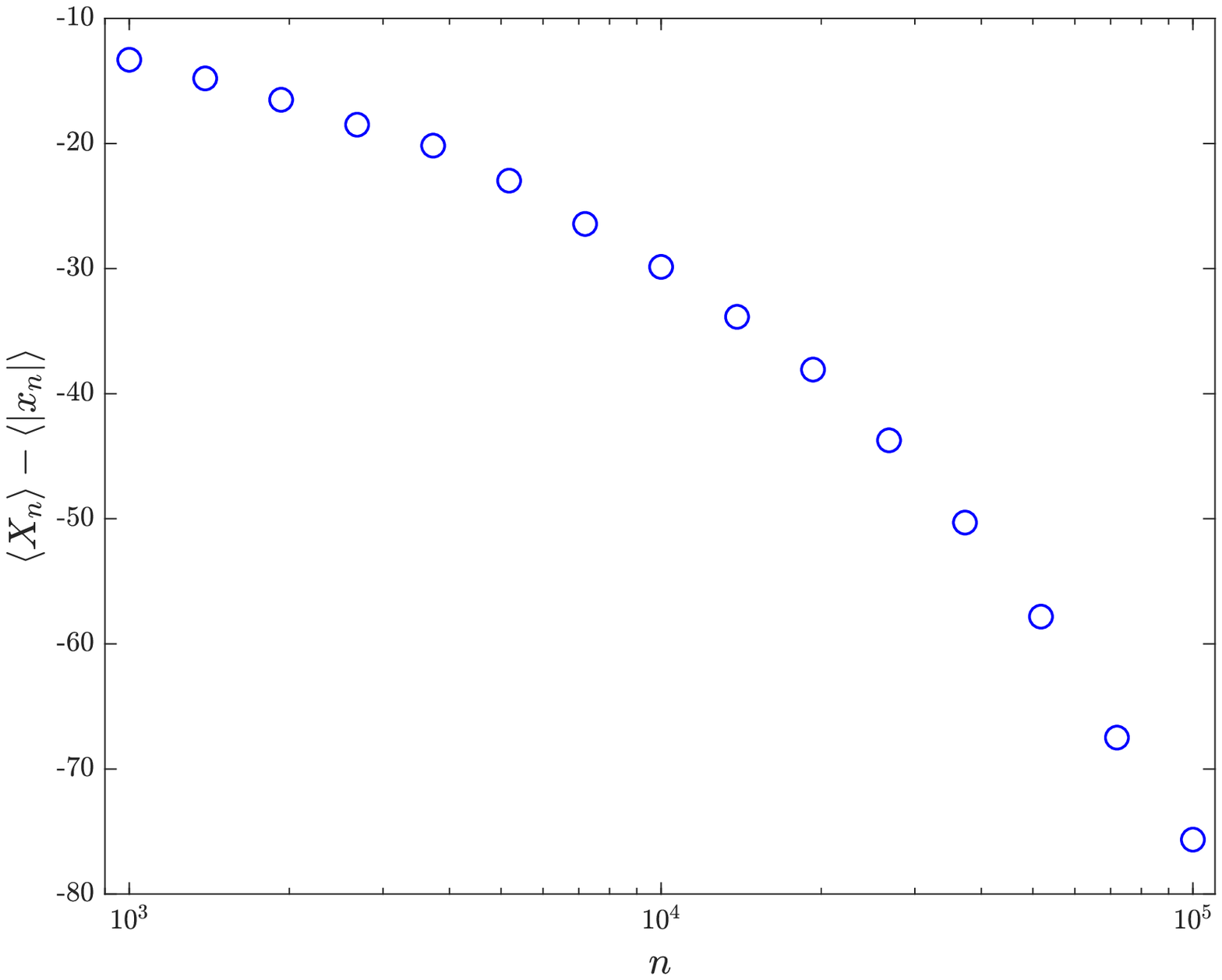} &
\includegraphics[width=.45\linewidth]{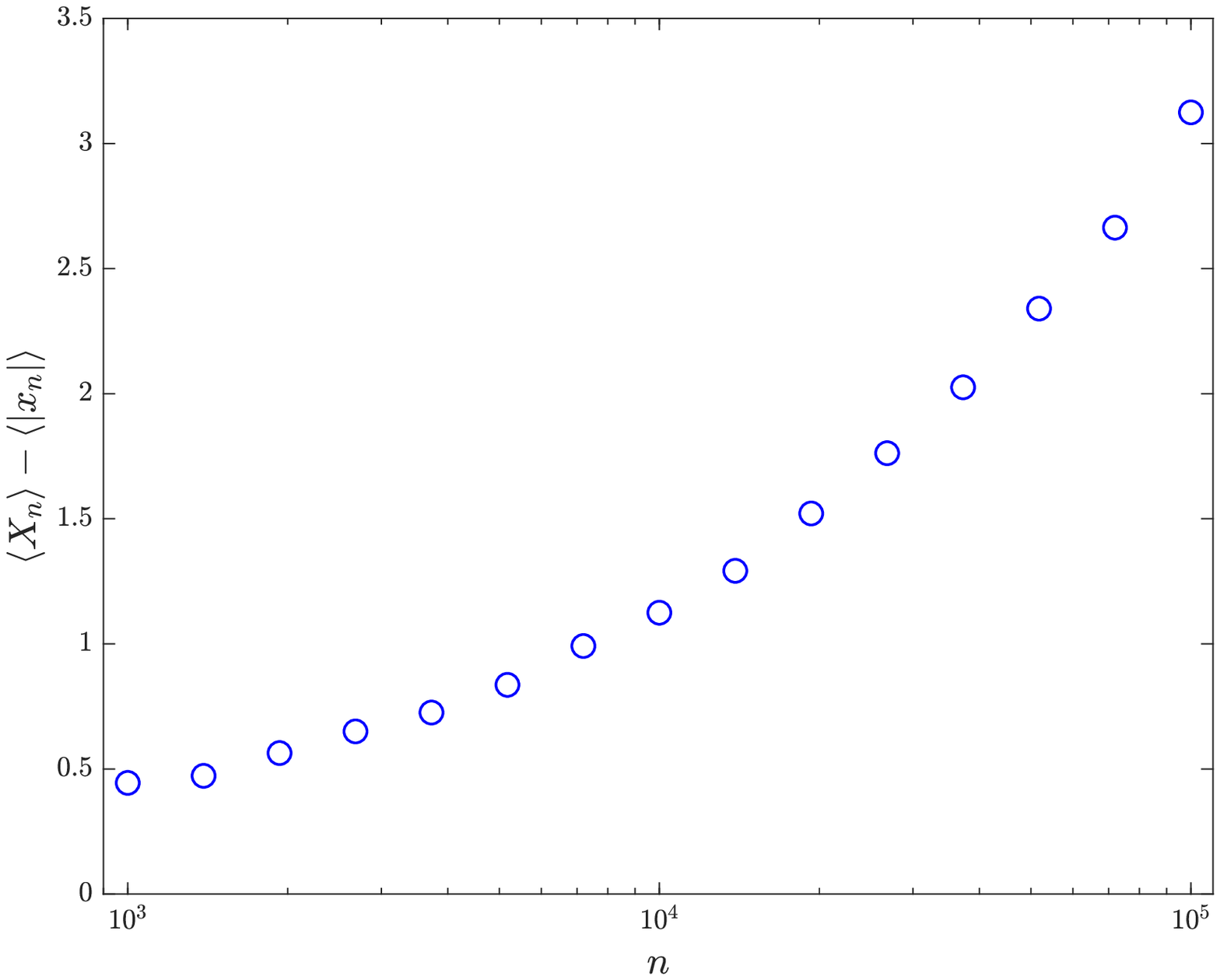} \\
\small (a) & \small (b) \\
\includegraphics[width=.45\linewidth]{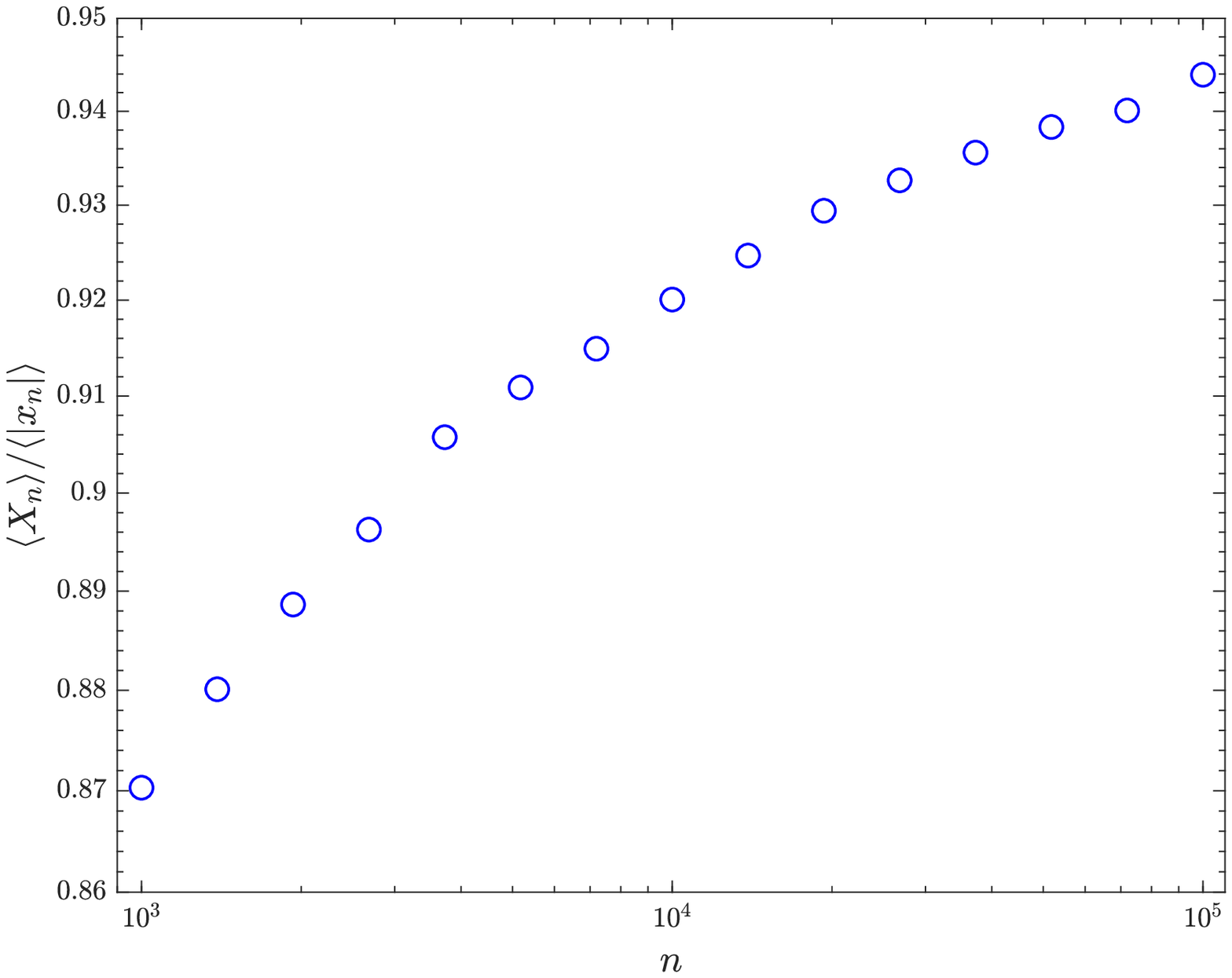} &
\includegraphics[width=.45\linewidth]{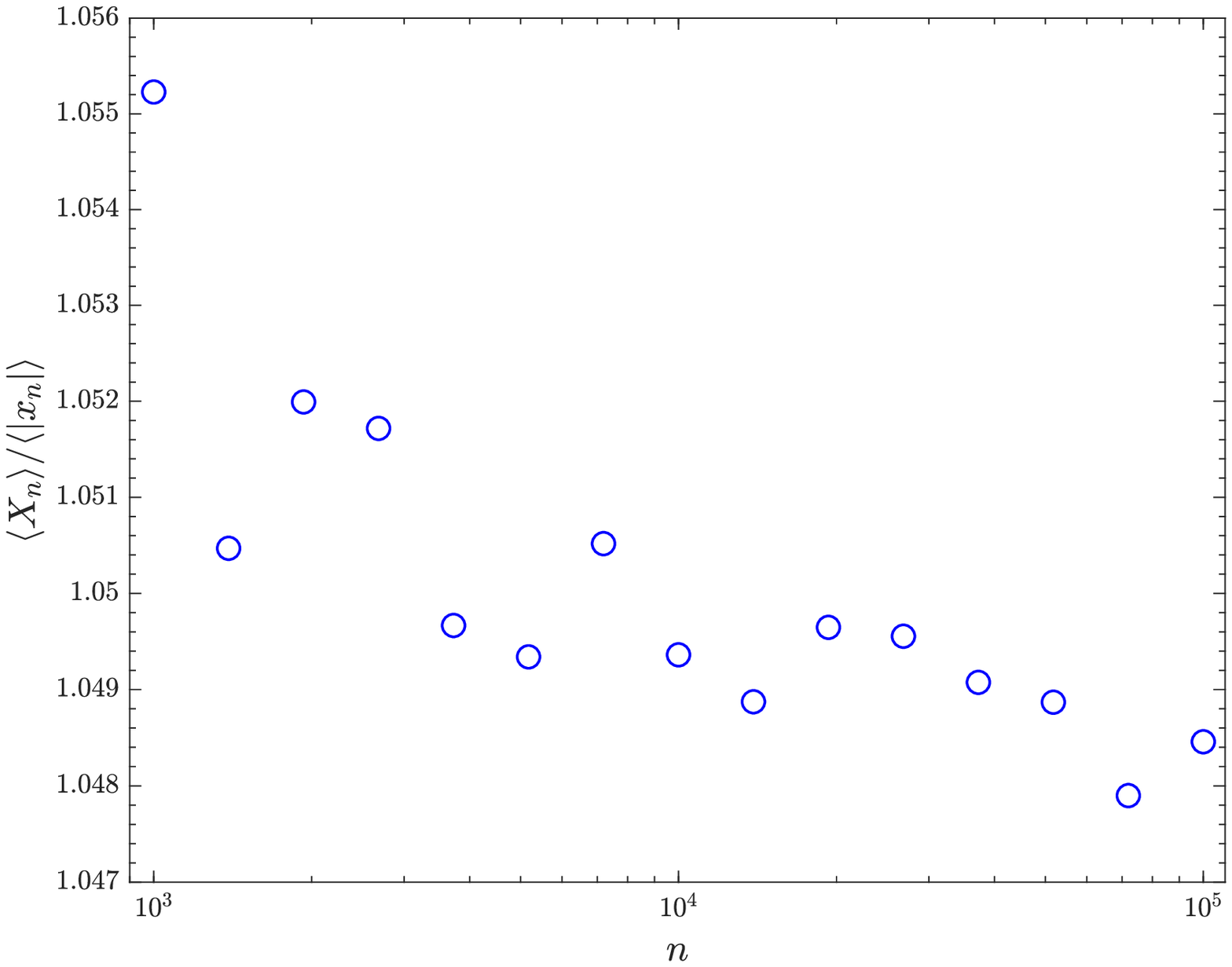} \\
\small (c) & \small (d)
\end{tabular}
\caption{Top (bottom): difference (ratio) between expected maximum $ \langle X_n \rangle $ and first moment $\langle |x_n| \rangle$ both for the superdiffusive, on the left, and subdiffusive model, on the right. Data are obtained by averaging $N=10^6$ walks composed of $n=10^5$ number of steps and for $\alpha=0.8$.}
\label{record:first_moment:diff-rat}
\end{figure*}

It is also interesting to investigate how the difference between $ \langle X_n \rangle$ and $\langle| x_n |\rangle$ behaves for large number of steps. For example, in a random walk with independent and identically distributed jumps such difference tends to a negative constant \cite{Com-Maj}. For both our models the difference diverges, and it is always negative for the superdiffusive model (fig. \ref{record:first_moment:diff-rat}a) and always positive for the subdiffusive one (fig. \ref{record:first_moment:diff-rat}b). The last check we made consists in plotting the ratio $ \langle X_n \rangle/\langle |x_n| \rangle$: we notice, as we could expect by the common form of asymptotic growth, that it tends to $1$ in both cases (figs. \ref{record:first_moment:diff-rat}c-\ref{record:first_moment:diff-rat}d).

\section{\label{s:5}Ageing}
As we already pointed out, to characterize the transport properties of a stochastic process the quantity usually considered is the mean square displacement. In most cases the measure of this observable is performed over a time corresponding to the duration of the process, viz. the measure and the process start at the same time $ t_0 $. However it may be worth analysing what happens if the observation time starts at a later instant $ t_a>t_0 $. If some dependence on $ t_a $ exists for the measured quantity, we say that the process exhibits {\it ageing}, and the instant $ t_a $ is called {\it ageing time} \cite{Kla-Sok}.

Consider the {\it aged ensemble-averaged} mean square displacement (AEMSD):
\begin{equation}\label{key}
\langle x^2\left(t;t_a\right)\rangle = \langle \left[x\left(t_a+t\right)-x\left(t_a\right)\right]^2\rangle.
\end{equation}
For a diffusion process with space-dependent diffusivity $ D(x)\sim |x|^{\gamma} $ such that the MSD grows as $ t^{\mu} $, with $ \mu=2/(2-\gamma) $, it is possible to show that the AEMSD depends explicitly on the ageing time \cite{Che-Chr-Met}:
\begin{equation}\label{eq:ageing:AEMSD}
\langle x^2\left(t;t_a\right)\rangle = \langle x^2(t)\rangle\cdot\mathcal{G}_\mu\left(\frac{t}{t_a}\right)
\end{equation}
where we defined the function $ \mathcal{G}_\mu(z) $ as:
\begin{equation}\label{key}
\mathcal{G}_\mu(z)=\frac{(1+z)^\mu-1}{z^\mu}.
\end{equation}
This shows that ageing is typical of anomalous diffusion processes, since in the limit $ \mu\to 1 $ the function $ \mathcal{G}_\mu (z)$ converges to the constant value $ 1 $, erasing the dependence on $ t_a $.

Ageing effects can also be studied when analysing the {\it time-averaged} mean squared displacement (TMSD):
\begin{equation}
\overline{\delta^2_T(\tau)}=\frac{1}{T-\tau}\int_0^{T-\tau}[x(t+\tau)-x(t)]^2 dt,
\label{Aging:TMSD:not_aging}
\end{equation}
where $T$ is the available process time length and $\tau$ is the lag time. This quantity is often considered in experiments, since it only requires the observation of a single trajectory for a sufficiently long time. In presence of ageing, the integration is performed starting from time $ t_a $ and so one can define the {\it aged time-averaged} mean square displacement (ATMSD):
\begin{equation}\label{key}
\overline{\delta^2_T(\tau;t_a)}=\frac{1}{T-\tau}\int_{t_a}^{t_a+T-\tau}[x(t+\tau)-x(t)]^2 dt.
\label{Aging:TMSD:aging}
\end{equation}
It is possible to evaluate the ATMSD by performing its ensemble average: interchanging the order of the averages one gets
\begin{equation}\label{key}
\langle\overline{\delta^2_T(\tau;t_a)}\rangle=\langle x^2(\tau)\rangle\cdot\frac{\tau}{T-\tau}\int_{t_a/\tau}^{(t_a+T)/\tau-1}\mathcal{G}_\mu\left(\frac{1}{z}\right)\mathrm{d}z
\end{equation}
and finally
\begin{equation}\label{key}
\langle\overline{\delta^2_T(\tau;t_a)}\rangle=\frac{\langle x^2(\tau)\rangle}{\tau^\mu}\cdot\frac{\mathcal{L}_{\mu+1}\left(T,t_a,\tau\right)}{T-\tau}
\end{equation}
where
\begin{equation}\label{key}
\mathcal{L}_\nu\left(T,t_a,\tau\right)=\frac{(t_a+T)^{\nu}-(t_a+\tau)^{\nu}-(t_a+T-\tau)^{\nu}+t_a^{\nu}}{\nu}.
\end{equation}

We notice that the ensemble average of the ATMSD is not equal to the AEMSD. Moreover, even in the absence of ageing time, in the limit $ T\gg\tau $ one gets \cite{Che-Chr-Met-II}:
\begin{equation}\label{eq:ageing:TMSD}
\langle\overline{\delta^2_T(\tau)}\rangle\sim\langle x^2(\tau)\rangle\cdot\left(\frac{\tau}{T}\right)^{1-\mu},
\end{equation}
meaning that the limit 
\begin{equation}
\lim_{T\to \infty} \overline{\delta^2_T(\tau)} = \langle x^2(\tau) \rangle
\end{equation}
does not hold. This feature of anomalous diffusion processes is referred to as \textit{weak ergodicity breaking} \cite{Che-Chr-Met, Che-Chr-Met-II, Che-Met, Met-Jeo-Che-Bar, Bur-Jeo-Met-Bar}. Finally we observe that in the limit $ T,T_a\gg\tau $ the ATMSD becomes
\begin{equation}\label{eq:ageing:ATMSD}
\langle\overline{\delta^2_T(\tau;t_a)}\rangle\sim\overline{\delta^2_T(\tau)}\cdot\mathcal{G}_\mu\left(\frac{T}{t_a}\right)
\end{equation}
so that, as for the AEMSD in eq. (\ref{eq:ageing:AEMSD}), the aged quantity is related to the unaged one by the factor $ \mathcal{G}_\mu(z) $.

The discussion made so far suggests that the averaged L\'evy-Lorentz gas exhibits both ageing effects and weak ergodicity breaking. Indeed the diffusion coefficients obtained for the averaged L\'evy-Lorentz gas and his localized version in the continuum limit are
\begin{equation}
D_{\alpha}(x)\sim\left\{
\begin{array}{ll}
| x |^{1-\alpha} &\textrm{averaged LL}\\
| x |^{\alpha-1} &\textrm{averaged-localized LL}
\end{array}
\right.
\end{equation}
Hence
\begin{equation}\label{key}
\langle x^2(t) \rangle \sim t^{\mu_\alpha}
\end{equation}
where 
\begin{equation}\label{Aging:K:LL}
\mu_\alpha=\left\{
\begin{array}{ll}
\frac{2}{1+\alpha},&\textrm{averaged LL}\\
\frac{2}{3-\alpha}&\textrm{averaged-localized LL}
\end{array}
\right.
\end{equation}
Therefore we expect to get 
\begin{equation}\label{Aging:MSD-TMSD:LL}
\begin{array}{r@{}l}
\langle\overline{\delta^2_T(\tau)}\rangle  &{}\sim \frac{\tau}{T^{1-\mu_\alpha}},\\
\langle\overline{\delta^2_T(\tau;t_a)}\rangle  &{}\sim  \frac{\tau}{T^{1-\mu_\alpha}}\cdot\mathcal{G}_{\mu_\alpha}\left(\frac{T}{t_a}\right).
\end{array}
\end{equation}

\begin{figure*}[h!]
\centering
\begin{tabular}{c @{\quad} c }
\includegraphics[width=.45\linewidth]{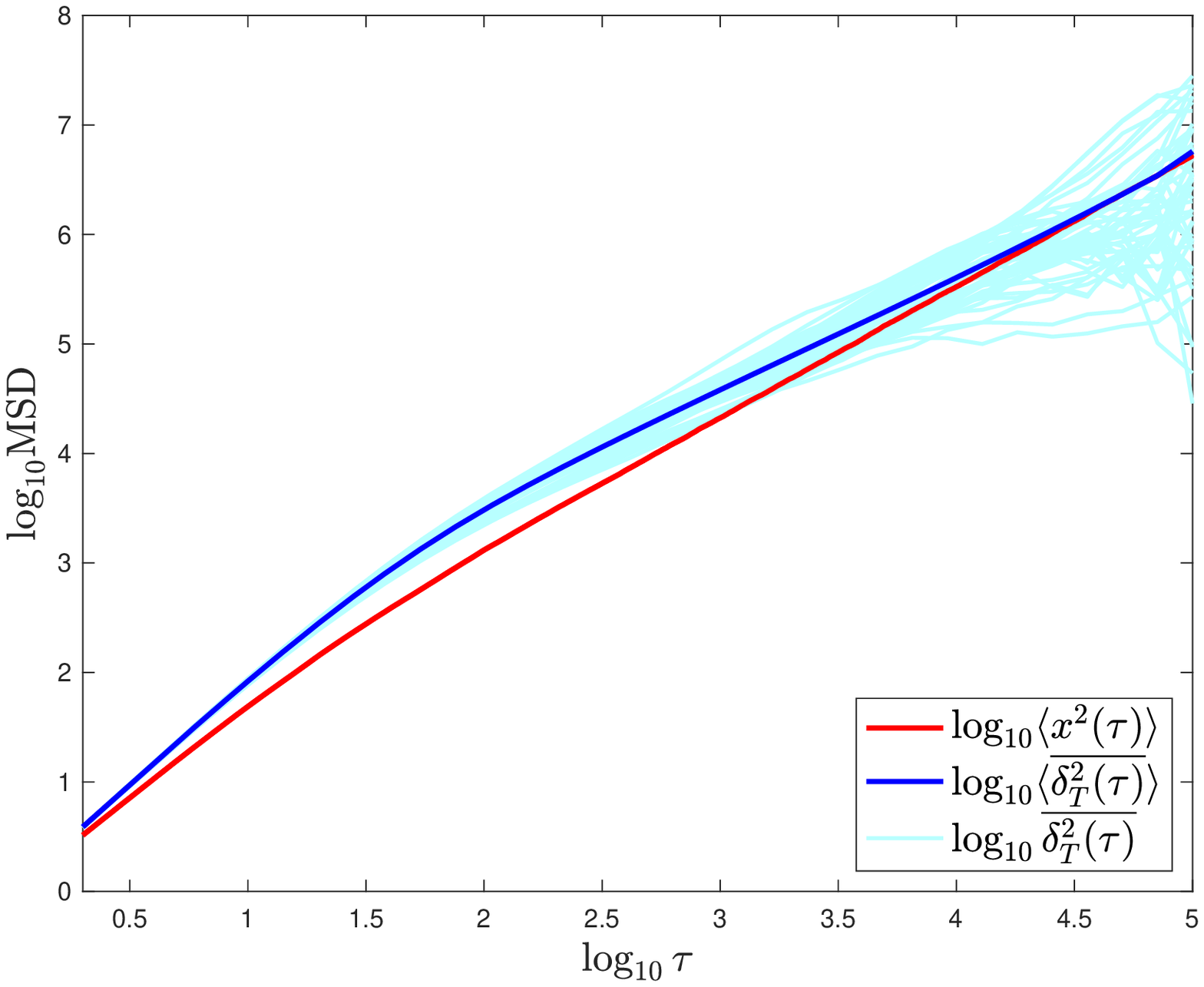} &
\includegraphics[width=.45\linewidth]{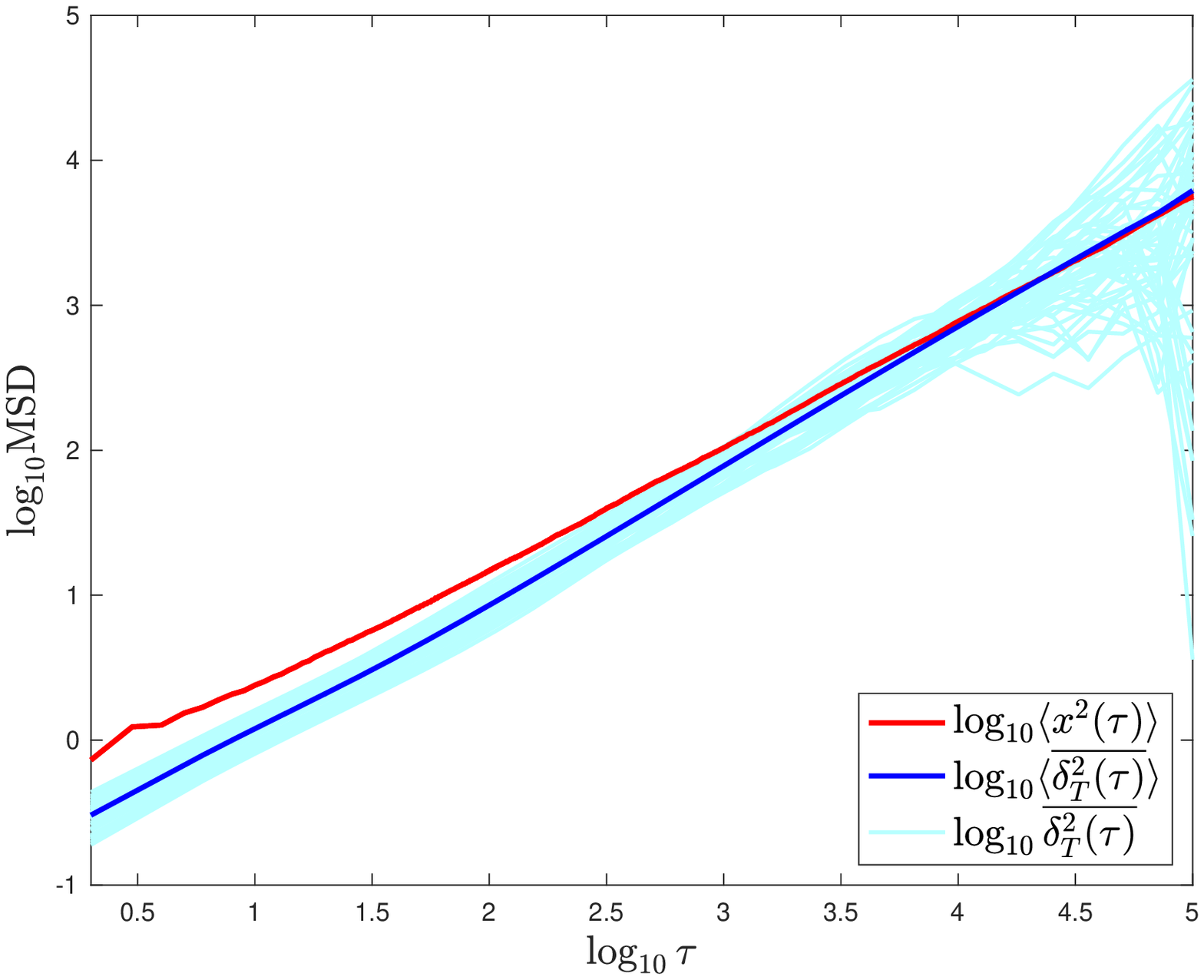} \\
\small (a) & \small (b) \\
\end{tabular}
\caption{Weak ergodicity breaking for both the Averaged LL gas (left) and its localized version (right) for $\alpha=0.7$. The red line shows the behaviour of $\langle x^2(\tau)\rangle$ depending on $\tau$, while the blue one represents the TMSD $\overline{\delta^2_T(\tau)}$ averaged over $N=10^4$ trajectories, of which only a few are plotted in cyan.  The time length of the process is $T=10^5$.}
\label{fig:weak:erg:breaking}
\end{figure*}

Fig. \ref{fig:weak:erg:breaking} shows our simulations confirming the weak ergodicity breaking for the Averaged L\'evy-Lorentz gas and its localized version. The TMSD is evaluated for a number $ N_{\mathrm{tr}}=10^4 $ of individual trajectories: the result varies from one single trajectory to another, meaning that the outcome of the TMSD is a random variable. This fact has already been observed for other systems for which ergodicity is broken, see \cite{He-Bur-Met-Bar} for further discussion. We notice that the average over the trajectories of the respective TMSD differs from the ensemble averaged MSD, and the two converge only for $\tau\sim T$, as expected from eq. (\ref{eq:ageing:TMSD}). Moreover the two systems display a difference in the fact that in the superdiffusive case the ensemble averaged MSD $ \langle x^2(\tau)\rangle $ is always smaller than $ \langle \overline{\delta^2_T(\tau)} \rangle $, while in the subdiffusive case one observes the opposite.

\begin{figure*}[h!]
\centering
\begin{tabular}{c @{\quad} c }
\includegraphics[width=.45\linewidth]{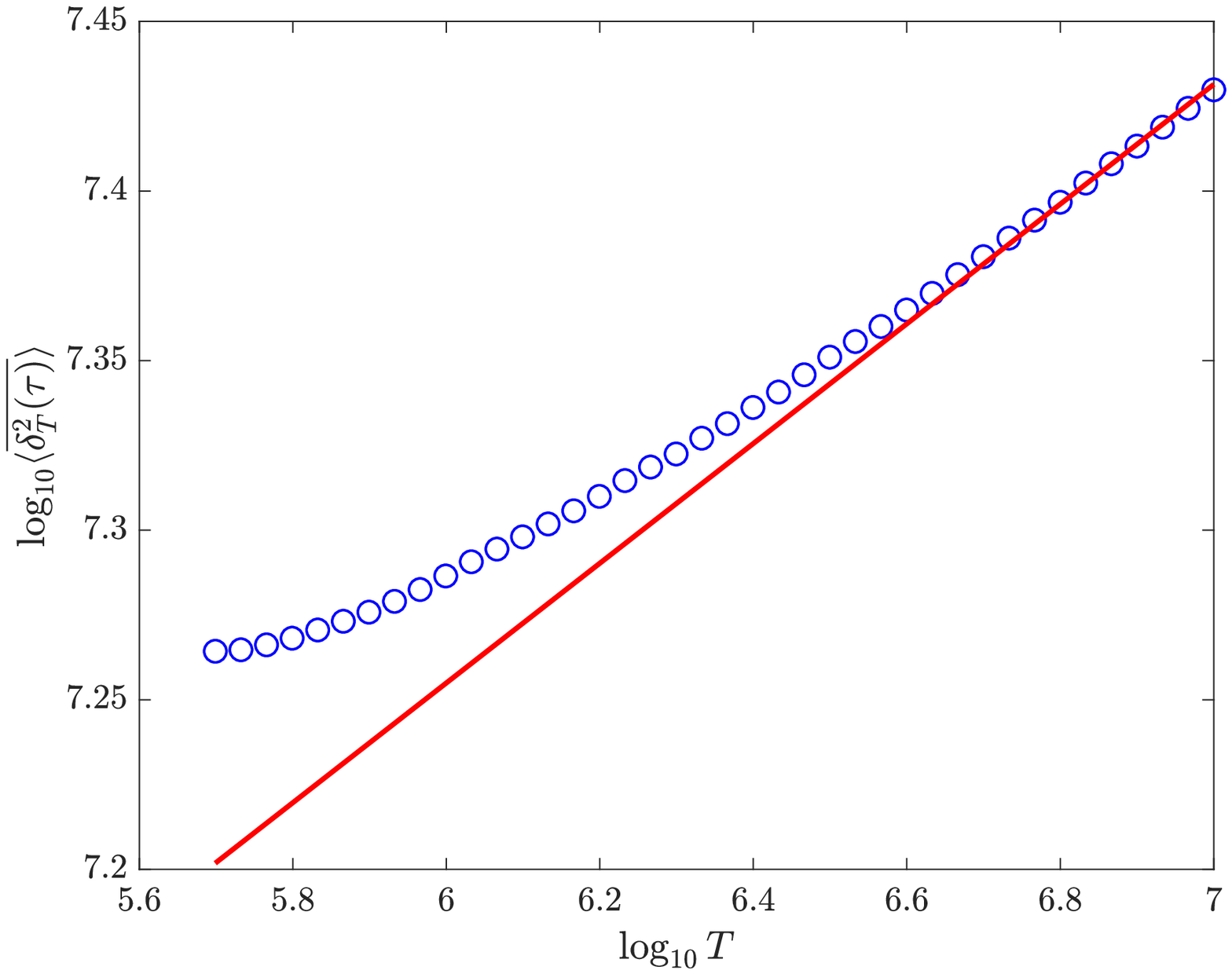} &
\includegraphics[width=.45\linewidth]{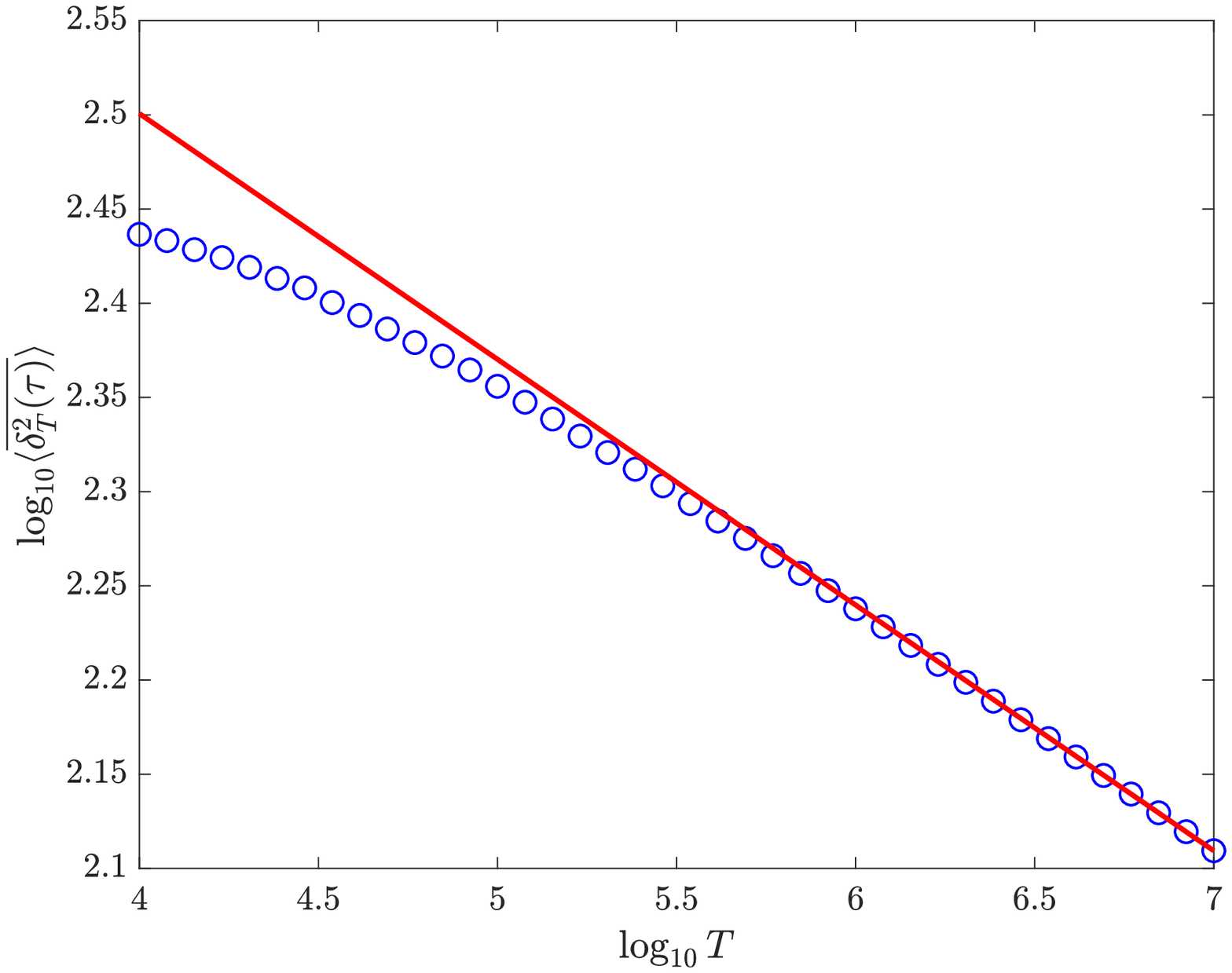} \\
\small (a) & \small (d) \\
\includegraphics[width=.45\linewidth]{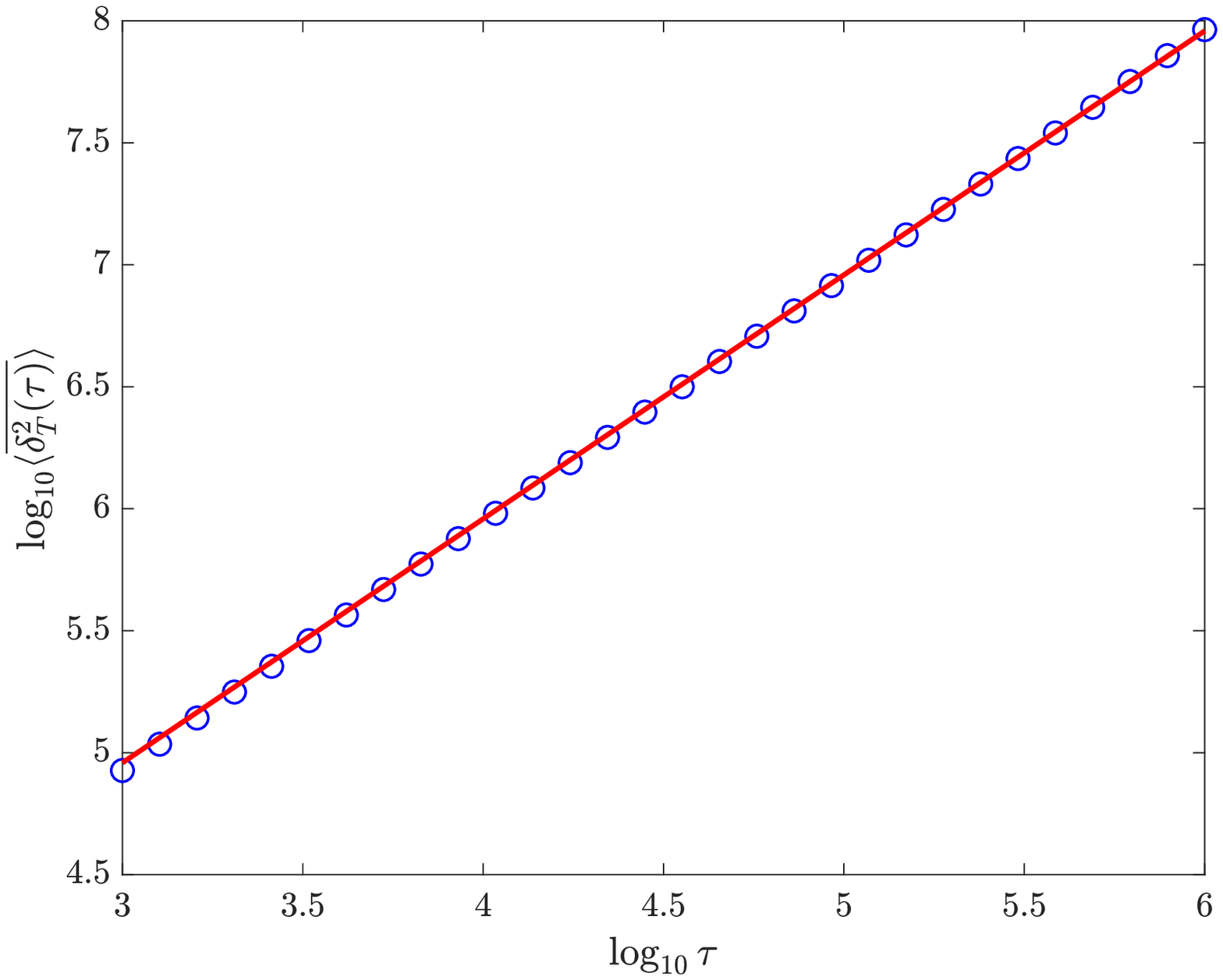} &
\includegraphics[width=.45\linewidth]{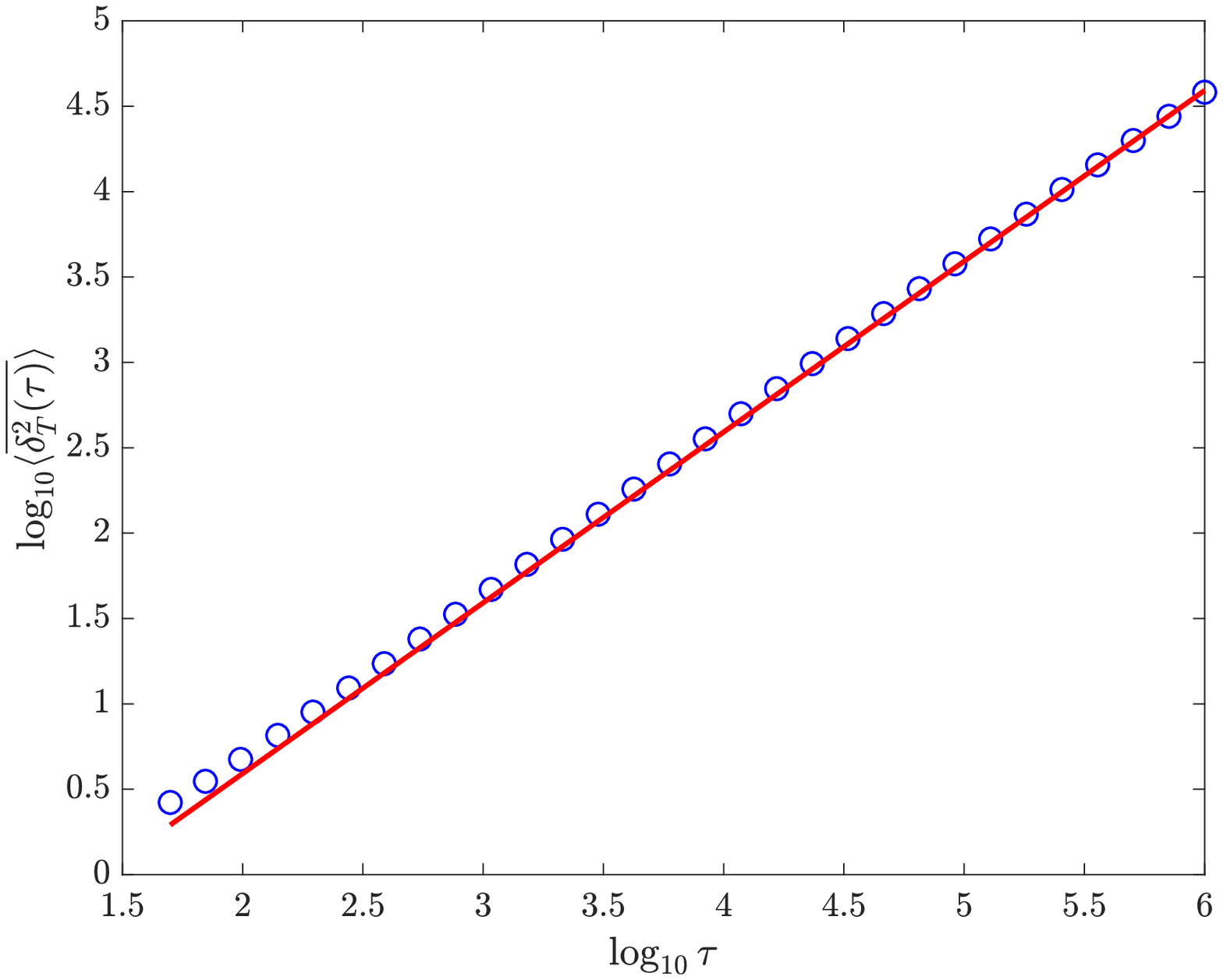} \\
\small (b) & \small (e)\\
\includegraphics[width=.45\linewidth]{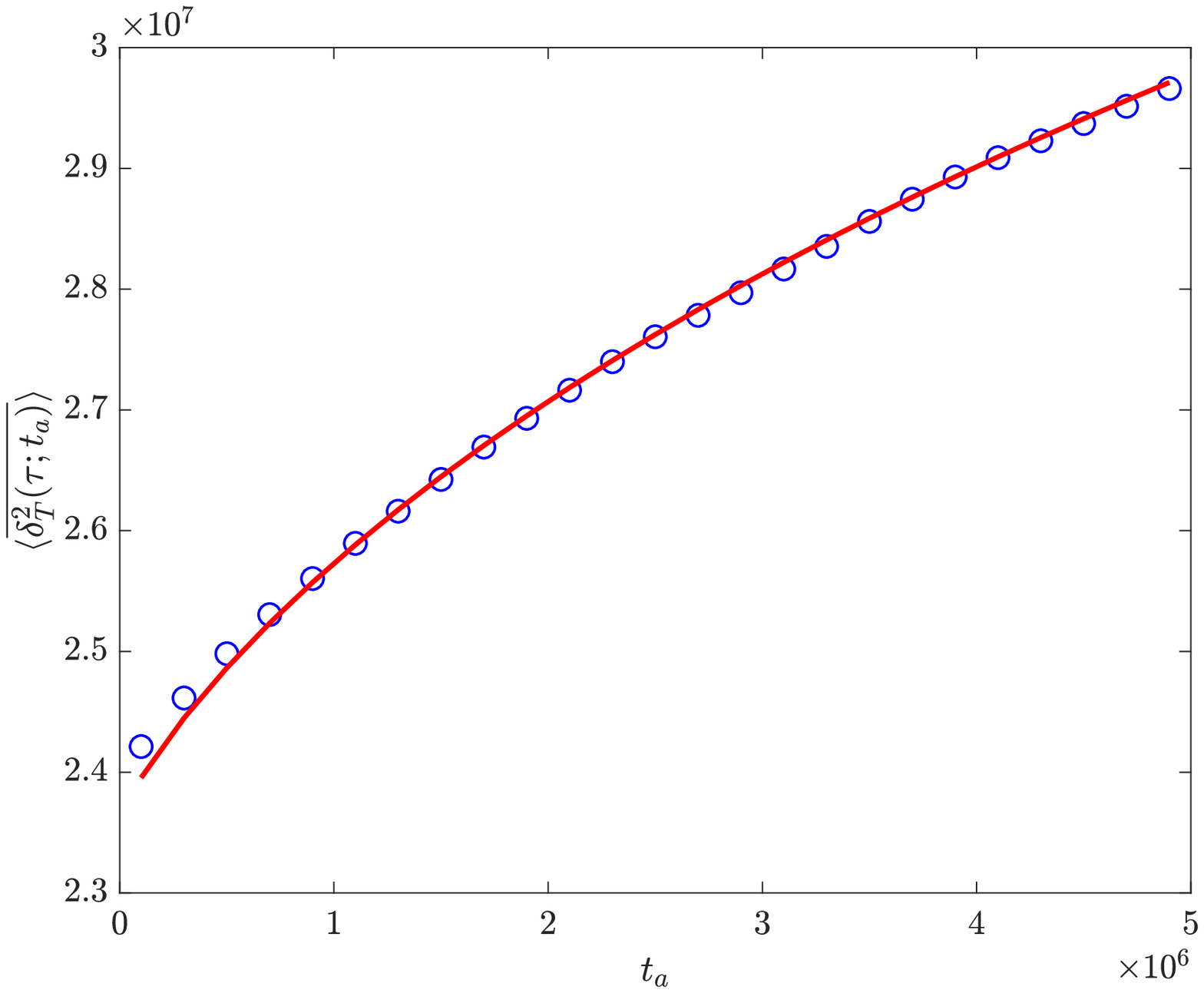} &
\includegraphics[width=.45\linewidth]{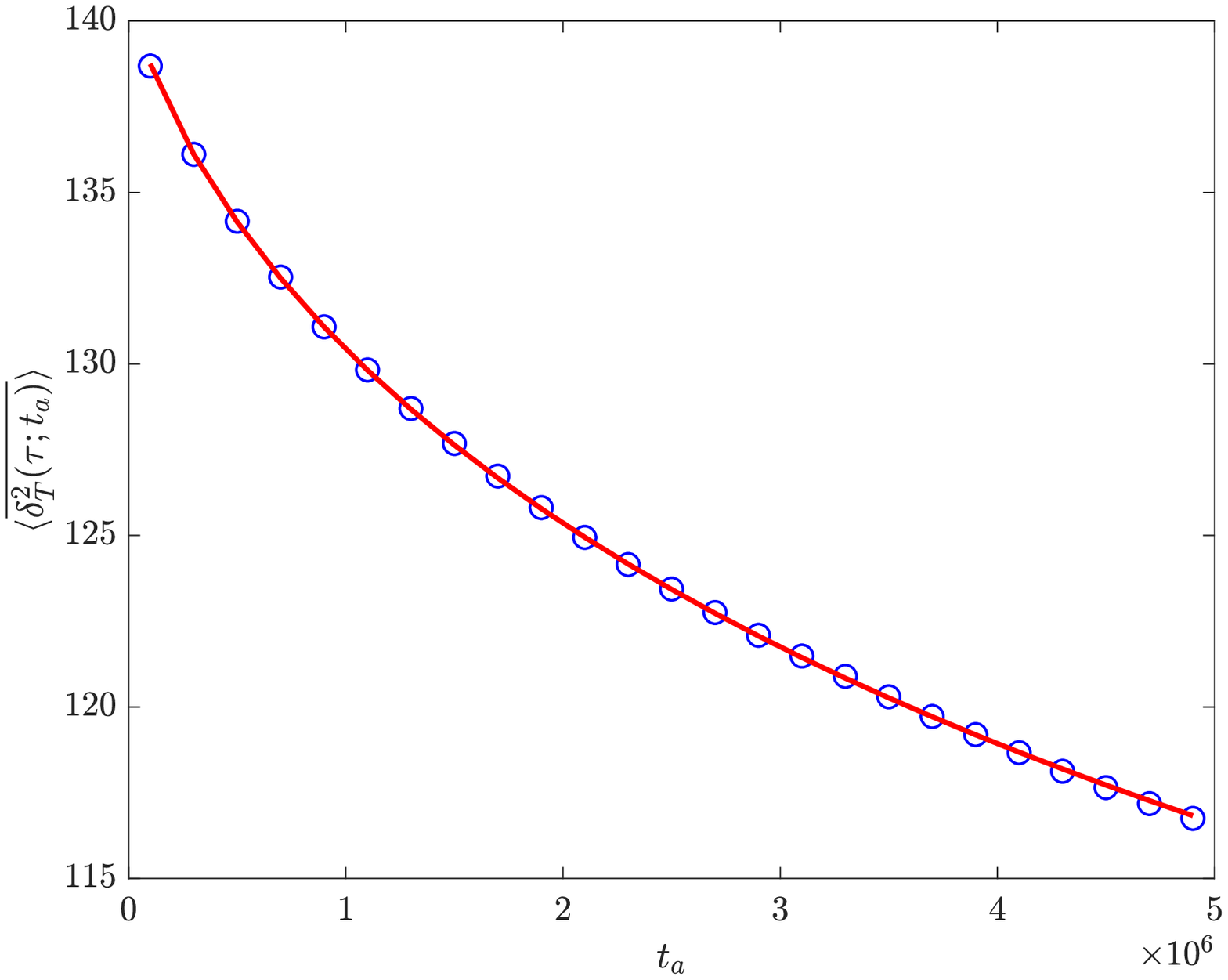} \\
\small (c) & \small (f)
\end{tabular}
\caption{Left (right): behaviour of the TMSD for the superdiffusive (subdiffusive) averaged L\'evy-Lorentz gas for $\alpha=0.7$. (a), (d): time averaged mean squared displacement $\langle\overline{\delta_T^2(\tau)}\rangle$ depending on  the process time length $T$ with $\tau=3\cdot 10^3$. (b), (e) $\langle\overline{\delta_T^2(\tau)}\rangle$ depending on  the lag time $\tau$ with $T=10^7$. (c), (f): ATMSD $\langle \overline{\delta^2_T(\tau;t_a)} \rangle$ depending on ageing $t_a$ with $\tau=3\cdot10^3$ and $T=5\cdot10^6$. The red lines show the asymptotic behaviours according to (\ref{Aging:MSD-TMSD:LL}). }
\label{Aging:fig:LL:sup:sub}
\end{figure*}

In fig. \ref{Aging:fig:LL:sup:sub}a (\ref{Aging:fig:LL:sup:sub}d) we present the results concerning the behaviour of $\langle\overline{\delta_T^2(\tau)}\rangle$ depending on $T$  with lag time $\tau=3\cdot 10^3$, for the superdiffusive (subdiffusive) case. The simulations results agree asymptotically with analytical ones. Figs. \ref{Aging:fig:LL:sup:sub}b and \ref{Aging:fig:LL:sup:sub}e show for both models the behaviour of $\langle\overline{\delta_T^2(\tau)}\rangle$ depending on $\tau$ with process time length $T=10^7$. The behaviour of the aged quantity $\langle\overline{\delta^2_T(\tau;t_a)}\rangle$ depending on $ t_a $ is shown in figures \ref{Aging:fig:LL:sup:sub}c and \ref{Aging:fig:LL:sup:sub}f, where we set $\tau=3\cdot 10^3$ and $T=5\cdot 10^6$.  The ageing time is taken of the order of the process time length. All the results are obtained for $ \alpha=0.7 $ by considering the averages over a number of trajectories $N_{\mathrm{tr}}=3\cdot 10^4$ for the subdiffusive case and $N_{\mathrm{tr}}=8\cdot 10^3$ for the subdiffusive one.

\section{\label{s:6}Conclusions}
We have considered the averaged version of the L\'evy-Lorentz gas, consisting in a non-homogeneous persistent random walk on a lattice with a power-law decay of the reflection probability. We have presented and analysed the localized model, which is trivial in its quenched version, since the motion is confined and there is not diffusion. In the related averaged version instead, the model is diffusive for $ 1<\alpha<2 $, while for $ 0<\alpha<1 $ the system displays weak anomalous diffusion. In particular we have subdiffusion, with an analytically computed MSD of the form $ \langle x_t^2\rangle \sim t^{2/(3-\alpha)} $.

Moreover, for both the localized subdiffusive model and the superdiffusive one we introduced in \cite{acor}, we presented  numerical results regarding the expected number of records and the expected value of the maximum of the random walk up to time $ n $, and discussed some numerical experiments proving that both models are affected by ageing and characterized by weak ergodicity breaking.

\section*{Acknowledgements}
The authors are very grateful to the referees for interesting and helpful comments. We acknowledge partial support by the research project PRIN 2017S35EHN\_007 ``Regular and stochastic behaviour in dynamical systems" of the Italian Ministry of Education and Research.

\end{document}